\documentclass[reprint,amsmath,amssymb,aps,prl,superscriptaddress]{revtex4-2}
\usepackage{graphicx}
\usepackage{dcolumn}
\usepackage{bm}
\usepackage{blindtext}
\usepackage{xcolor}
\usepackage{tabularx}
\usepackage{booktabs}
\usepackage{geometry}
\usepackage{array}
\usepackage{appendix}
\usepackage{placeins}
\usepackage{amsmath}
\usepackage{hyperref}

\UseRawInputEncoding

\newif\ifarxiv
\arxivtrue 

\begin{document}
\title{Phase Stability of Lead Phosphate Apatite Pb$_{10-x}$Cu$_{x}$(PO$_{4}$)$_{6}$O, Pb$_{10-x}$Cu$_{x}$(PO$_{4}$)$_{6}$(OH)$_{2}$, and Pb$_{8}$Cu$_{2}$(PO$_{4}$)$_{6}$}

\author{Jiahong Shen}
\author{Dale Gaines II} 
\author{Shima Shahabfar} 
\author{Zhi Li}
\author{Dohun Kang}
\author{Sean Griesemer}
\author{Adolfo Salgado-Casanova}
\author{Tzu-chen Liu}
\author{Chang-Ti Chou}
\affiliation{Department of Materials Science and Engineering, Northwestern University, Evanston, IL 60208, USA} 
\author{Yi Xia}
\affiliation{Department of Mechanical and Materials Engineering, Portland State University, Portland, OR 97201, USA}
\author{Chris Wolverton}
\affiliation{Department of Materials Science and Engineering, Northwestern University, Evanston, IL 60208, USA} 

\date{August 14th, 2023}

\begin{abstract}
Recently, Cu-substituted lead apatite LK-99 was reported to have room-temperature ambient-pressure superconductivity. Here we utilize density functional theory (DFT) total energy and harmonic phonon calculations to investigate the thermodynamic and dynamic stability of two lead phosphate apatites in their pure and Cu-substituted structures. Though Pb$_{10}$(PO$_4$)$_6$O and Pb$_{10}$(PO$_4$)$_6$(OH)$_2$ are found to be thermodynamically stable (i.e., on the T=0K ground state convex hull), their Cu-substituted counterparts are above the convex hull. Harmonic phonon calculations reveal dynamic instabilities in all four of these structures. Oxygen vacancy formation energies demonstrate that the addition of Cu dopant substituting for Pb increases the likelihood of the formation of oxygen vacancies on the anion site. We propose a new possible phase in this system, Pb$_8$Cu$_2$(PO$_4$)$_6$, where two monovalent Cu atoms are substituted for two Pb(1) atoms and the anion oxygen is removed. We also propose several reaction pathways for Pb$_9$Cu(PO$_4$)$_6$O and Pb$_8$Cu$_2$(PO$_4$)$_6$, and found that both of these two structures are likely to be synthesized under a 1:1 ratio of reactants Pb$_2$SO$_5$ and Cu$_3$P. Our work provides a thorough foundation for the thermodynamic and dynamic stabilities of LK-99 related compounds and we propose several possible novel synthesis reaction pathways and a new predicted structure for future studies. 
\end{abstract}

\maketitle
\section{Introduction}
Superconductors exhibit zero electrical resistance when under specific conditions and this unique feature opens the door to a wide array of technological advancements that have the potential to address some of the most pressing challenges in energy~\cite{Holmes2013}, transportation~\cite{Rote}, medicine~\cite{Ali2022}, and beyond. The first superconductor was mercury at 4.19 K found by Onnes in 1911~\cite{onnes1991}. Since that time, a large variety of superconducting materials have been found. Low temperature ($<$77 K, which is the liquid nitrogen boiling point) can trigger superconducting behavior~\cite{Einsenstein1954}, but such low temperatures pose a formidable challenge for all applications. Consequently, the pursuit of high-temperature ($>$77 K) superconductors~\cite{Bednorz1986,Subramanian1988} or even room-temperature superconductors (RTSC) has become a paramount objective in the field. High pressure experiments can also lead to superconductivity at elevated temperatures~\cite{Drozdov2015,Drozdov22019,Sun2019}. But, finding a material which exhibits superconductivity under ambient conditions remains a grand challenge.

In a recent report (at the time of this writing, still unverified), Lee {\it{et\,al.}} reported that a Cu-doped lead apatite material named LK-99~\cite{Sukbae23,Sukbaelee} exhibited superconductivity at room temperature and under ambient pressure conditions. The authors argued that the Cu-doping and its resulting structural distortion lead to the superconducting behavior. LK-99$\rq$s parent structure, Lead phosphate apatite, contains a mixed polyanionic framework, where phosphorus atoms are four-fold coordinated with oxygen to form (PO$_{4}$)$_{x}$ tetrahedra, constituting the building blocks of lead phosphate structures. These tetrahedra can be flexibly arranged in various ways, facilitating the assembly of diverse P$_{x}$O$_{y}$ groups, resulting in the formation of intricate motifs such as rings, chains~\cite{Zhao2014}, and isolated dimers~\cite{Dong2016}. Moreover, lead phosphates can adopt multiple different stoichiometries and polymorphs, which often exhibit near degeneracies and metastabilities, resulting in a complex phase space.  

Recent Density Functional Theory (DFT) studies have examined some of the properties and electronic structure of various family members of lead phosphates, including LK-99. The presence of a flat band in the electronic band structure of lead apatites was determined, and its strongly correlated nature was hypothesized to play a role in the reported superconducting behavior in LK-99~\cite{Grifin2023,Rafal2023}.  The energetics of different magnetic orderings was studied by Cabezas-Escares {\it{et\,al.}}~\cite{Cabez23} and they showed that  ferromagnetic and antiferromagnetic configurations are practically degenerated in the case of Pb$_{10}$Cu$_{x}$(PO$_{4}$)$_{6}$O. However, there are many open questions regarding the thermodynamic, dynamic, and phase stabilities of LK-99 and related compounds in their doped and undoped forms.  And, the synthesis reaction pathways and energetics that favor the formation of Cu doping in lead apatites is also not well understood. In this paper, we study the thermodynamic and dynamic stability of Pb$_{10-x}$Cu$_{x}$(PO$_{4}$)$_{6}$O and Pb$_{10-x}$Cu$_{x}$(PO$_{4}$)$_{6}$(OH)$_{2}$ where x = 0 or 1, and explore their electronic and phonon band structures. We also propose a new predicted phase, Pb$_{8}$Cu$_{2}$(PO$_{4}$)$_{6}$  as a possible anion-free apatite and compare it with other structures. Finally, we will construct a reaction convex hull to evaluate various reaction pathways to synthesize LK-99 and propose an energetically promising reaction pathway.

\section{Methods}
\subsection{DFT Calculations}
All calculations were performed with VASP~\cite{VASP1,VASP2} using the PBE exchange-correlation functional~\cite{PBE}, with an energy cutoff of 520 eV, and gamma centered $k$-points grids with a $k$-spacing of 0.15 {\AA}$^{-1}$. Structural relaxation was performed with an energy convergence threshold of 10$^{-8}$ eV and a force convergence threshold of 10$^{-3}$ eV/{\AA}. The spin-polarized electronic structure was calculated using a fully relaxed crystal structure. The inclusion of electron-electron repulsion arising from localized Cu $d$-orbitals was achieved through the implementation of Dudarev's GGA+U approach~\cite{GGAU}. The U-J value was set at 4 eV, consistent with our prior configuration within the Open Quantum Materials Database (OQMD)~\cite{OQMD_2,OQMD_1}. Phonon calculations were performed using supercells of size 1$\times$1$\times$2 through the finite displacement method implemented in phonopy~\cite{Phonopy} without the inclusion of spin or DFT+U. Phonon band structures in the main text are truncated at 15 THz to show detail in the lower energy modes, but full phonon band structures are included in the Supplemental Information.

\subsection{Therodynamic Stability}
The thermodynamic stability of a compound was determined by convex hull constructions as implemented in the OQMD. The details can be found in the Methods section in J. Shen {\it{et\,al.}}~\cite{Jiahong2021}. In addition to the convex hull distance, another useful energetic metric is the decomposition energy, which can be calculated to represent the degree of stability of stable phases as follows:
$\Delta$H$_{\text{decomp}}$ = $\Delta$H$'_{\text{hull}}$ $-$ $\Delta$H$_{\text{f}}$.

In this equation, $\Delta$H$_{\text{decomp}}$ represents the decomposition energy of the compound, $\Delta$H$_{\text{f}}$ denotes the formation enthalpy of the compound, and $\Delta$H$_{\text{hull}}$ is the energy of the convex hull constructed without any compounds at that particular composition. In contrast to the convex hull distance $\Delta$H$_{\text{stability}}$, a higher value of $\Delta$H$_{\text{decomp}}$ means a structure is more stable, lying deeper in the convex hull.

\begin{figure*}[htp!]
\includegraphics[width=0.9\textwidth]{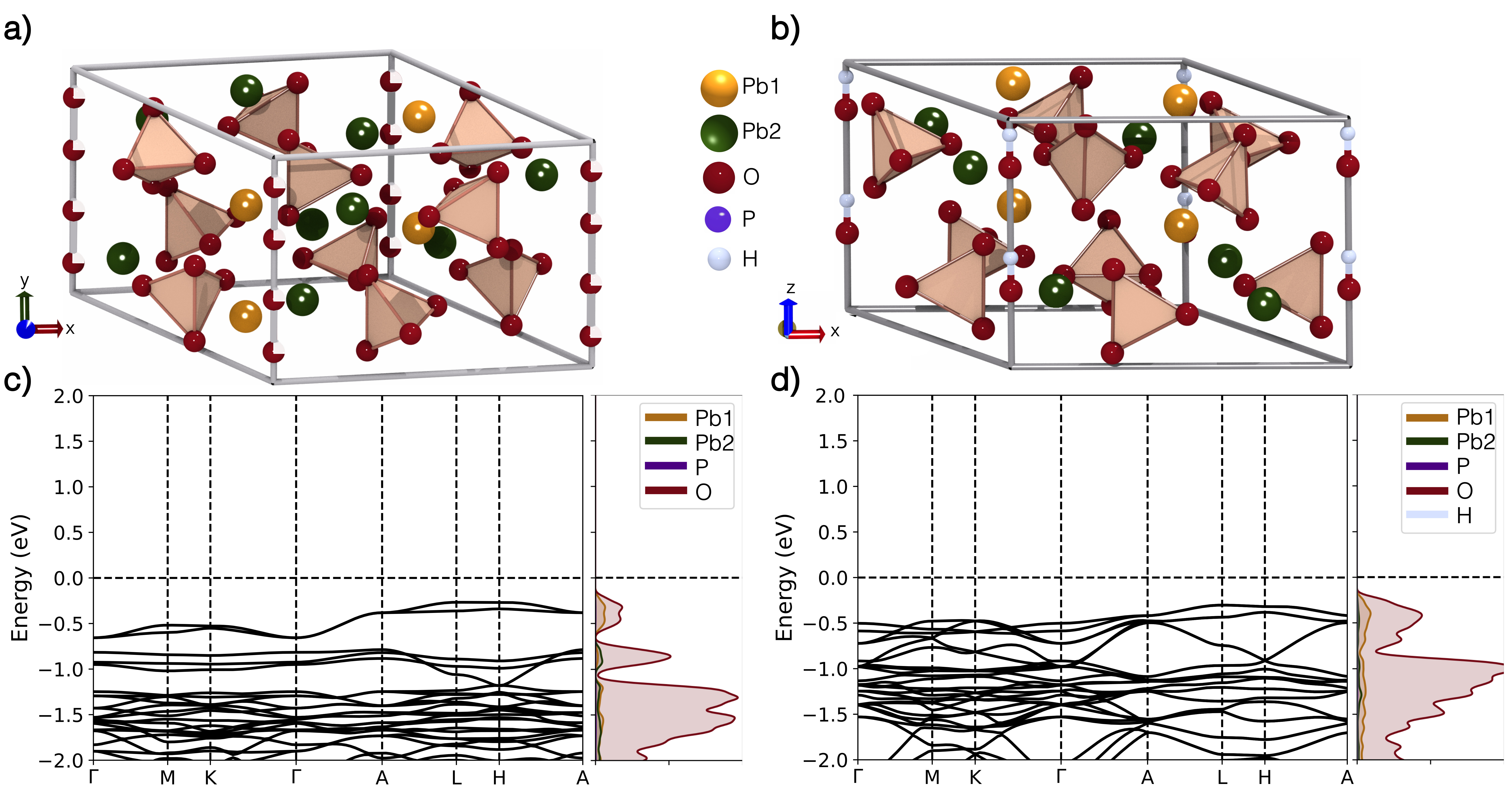}
\caption{a,b) The crystal structure of Pb$_{10}$(PO$_{4}$)$_{6}$O 1/4 partially occupied oxygen sites and Pb$_{10}$(PO$_{4}$)$_{6}$(OH)$_{2}$. c,d) band structure with atom-projected density of states of Pb$_{10}$(PO$_{4}$)$_{6}$O (one oxygen site fully occupied) and Pb$_{10}$(PO$_{4}$)$_{6}$(OH)$_{2}$.} 
\label{fig1}
\end{figure*}

\section{Results and Discussions}
LK-99 was reported as a Cu-substituted lead phosphate apatite. In this work, we studied both the LK-99 synthesized in vacuum Pb$_{10-x}$Cu$_{x}$(PO$_{4}$)$_{6}$O and that in the atmosphere Pb$_{10-x}$Cu$_{x}$(PO$_{4}$)$_{6}$(OH)$_{2}$. The Cu substitution ratio was reported as 0.9 $<$ x $<$ 1.1, and here we chose x = 0 or 1 for simplicity. For completeness of this work and to search for the structures that may better represent LK-99, we also consider the Pb$_{8}$Cu$_{2}$(PO$_{4}$)$_{6}$ with a monovalent Cu and without the extra oxygen atom. Our results were categorized into five subsections: 1. Lead pure Pb$_{10}$(PO$_{4}$)$_{6}$O, 2. Cu-substituted Pb$_{9}$Cu(PO$_{4}$)$_{6}$O, 3. Lead pure Pb$_{10}$(PO$_{4}$)$_{6}$(OH)$_{2}$, 4. Cu-substituted Pb$_9$Cu(PO$_4$)$_6$(OH)$_2$, and 5. Pb$_{8}$Cu$_{2}$(PO$_{4}$)$_{6}$.

\begin{figure*}[htp!]
\includegraphics[width=0.9\textwidth]{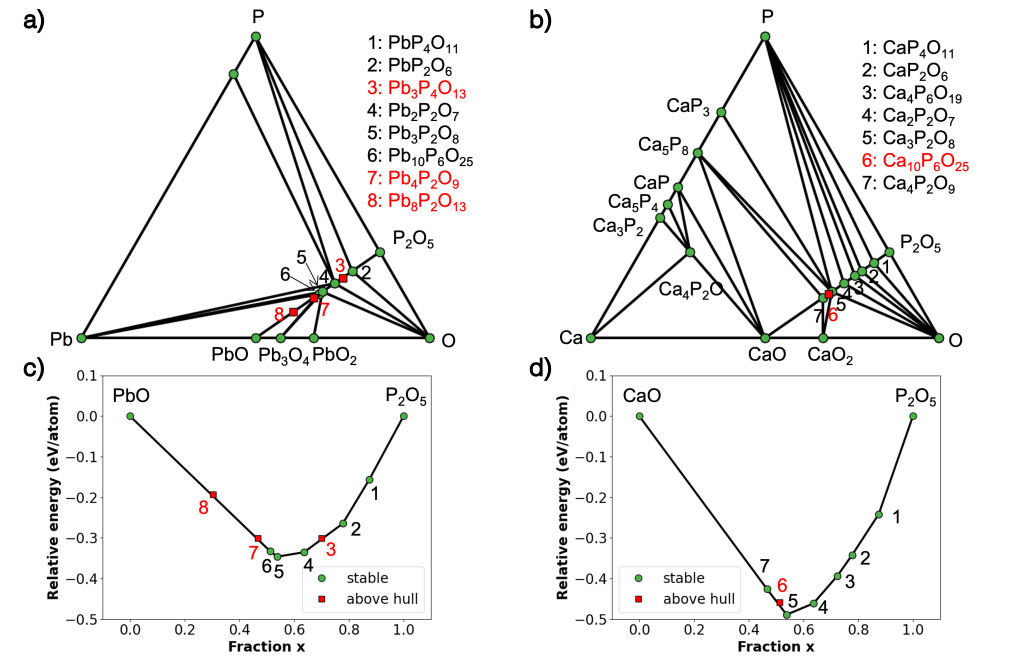}
\caption{Phase diagram of a) Pb--P--O, and b) Ca--P--O ternary phase space, and isopleth convex hull of c) PbO-P$_{2}$O$_{5}$, and d) CaO–-P$_{2}$O$_{5}$. The green circles and red squares indicate the phases that are stable (on the hull) or unstable (above the hull). PbP$_{4}$O$_{11}$ is a new stable compound discovered in this work.} 
\label{fig2}
\end{figure*}

\subsection{1. Pb$_{10}$(PO$_{4}$)$_{6}$O}
The crystal structure of Pb$_{10}$(PO$_{4}$)$_{6}$O is shown in Figure~\ref{fig1}(a). This is the parent structure of LK-99 and has a hexagonal P6$_{3}$/m  (176) space group. There are two inequivalent Pb atoms in the unit cell of Pb$_{10}$(PO$_{4}$)$_{6}$O based on the Wyckoff positions where Pb(1) occupies site 4f while Pb(2) occupies site 6h. There are also four equivalent O sites, which are not part of the structure's six PO$_{4}$ tetrahedra, that are 1/4 partially occupied. Liang Si {\it{et\,al.}} investigated the occupation of these O sites and found that they are energetically equivalent. Furthermore, after testing seven different O configurations in a 2$\times$2$\times$1 supercell of Pb$_{10}$(PO$_{4}$)$_{6}$O, they concluded that at room temperature and above, these O sites will exhibit some disorder in their occupations~\cite{Liang23}. As a result, we simply chose one site from the four partially occupied sites to put one oxygen atom and remove the other three. This results in a breaking of the symmetry and the resulting structure has a trigonal P3 (143) space group. The calculated electronic structure is shown in Figure~\ref{fig1}(c), suggesting that Pb$_{10}$(PO$_{4}$)$_{6}$O is a semiconductor with a wide bandgap of 2.77 eV. We can observe two flat bands (dispersion $\sim$~0.4 eV) right below the Fermi level. These bands originate from the 6$s^2$ lone-pair states of Pb(2) and the O 2$p$ states. The results are in good agreement with previous DFT works~\cite{Cabez23,Junwen2023,Liang23}.

\begin{table*}[htp!]
  \centering
  \renewcommand{\arraystretch}{1.2} 
\begin{tabularx}{\textwidth}{ |c|c|c|c|X|}

\hline
\textbf{Compound} & \textbf{Space Group} & \textbf{Stability(eV)} & \textbf{Ref.} & \textbf{Isotructural Compounds in ICSD} \\
\hline
Pb$_{2}$P$_{2}$O$_{7}$  & P-1 & 0 & \cite{Mullica1986} & Sn$_{2}$P$_{2}$O$_{7}$, V$_{2}$Cu$_{2}$O$_{7}$, Sr$_{2}$V$_{2}$O$_{7}$, Ca$_{2}$V$_{2}$O$_{7}$ ,Mg$_{2}$V$_{2}$O$_{7}$,Cd$_{2}$P$_{2}$O$_{7}$, Ba$_{2}$V$_{2}$O$_{7}$, Hg$_{2}$P$_{2}$O$_{7}$, Mn$_{2}$V$_{2}$O$_{7}$ \\
\hline
Pb$_{3}$P$_{2}$O$_{8}$ & C2/c & 0 & \cite{belokoneva2001} & Eu$_{3}$Ga$_{2}$, Zn$_{3}$Ta$_{2}$O$_{8}$, Zn$_{3}$Nb$_{2}$O$_{8}$ \\
\hline
PbP$_{2}$O$_{6}$ & $P2_{1}/c$ & 0 & \cite{Jost1964} & BeP$_{2}$O$_{6}$, CaI$_{2}$O$_{6}$, BaP$_{2}$O$_{6}$, SrNb$_{2}$O$_{6}$, CuSb$_{2}$O$_{6}$, BaNb$_{2}$O$_{6}$, Ta$_{2}$CrO$_{6}$, CaP$_{2}$O$_{6}$, CaNb$_{2}$O$_{6}$, EuNb$_{2}$O$_{6}$, SrP$_{2}$O$_{6}$, CuSb$_{2}$O$_{6}$  \\
\hline
Pb$_{10}$P$_{6}$O$_{25}$ & P6$_{3}$/m & 0 & \cite{Sukbae2023} & Sr$_{10}$P$_6$O$_25$, La$_{10}$Ge$_6$O$_{25}$\\
\hline
Pb$_{4}$P$_{2}$O$_{9}$ & $P2_{1}/c$ & 0.002 & \cite{Krivovichev2003} & Fe$_{4}$P$_{2}$O$_{9}$, Zn$_{4}$AS$_{2}$O$_{9}$, Sn$_{4}$P$_{2}$O$_{9}$, Ca$_{4}$Nb$_{2}$O$_{9}$ \\
\hline
Pb$_{3}$P$_{4}$O$_{13}$ & P-1 & 0.003 & \cite{Averbuch1987} & Ba$_{3}$P$_{4}$O$_{13}$, Sr$_{3}$P$_{4}$O$_{13}$ \\
\hline
Pb$_{8}$P$_{2}$O$_{13}$ & C2/m & 0.005 & \cite{Krivovichev2003} & Pb$_{8}$As$_{2}$O$_{13}$ \\
  \hline
  \hline
\multicolumn{5}{|c|}{\textbf{Hypothetical Compounds Created from Elemental Substitution}} \\
\hline
PbP$_{4}$O$_{11}$ & Aba2 & 0 & -- & CaP$_{4}$O$_{11}$ \\
\hline
Pb$_{4}$P$_{6}$O$_{19}$ & P-1 & 0.012 & -- & Ca$_{4}$P$_{6}$O$_{19}$ \\
\hline
\end{tabularx}
    \caption{Reported ternary compounds within 5 meV/atom the P$_{2}$O$_{5}$--PbO convex hull, corresponding space groups, and isostructural compounds in the ICSD. Last two rows are hypothetical compounds calculated in this work.}
    \label{tab:1}
\end{table*}

We next turn to the thermodynamic stability (defined in Methods) in the Pb--P--O system, which is an important factor to infer the synthesizability of a compound~\cite{Sunn2016,Jiahong2021}. The phase diagram of the Pb--P--O ternary phase space is calculated through the OQMD and is shown in Figure~\ref{fig2}(a). Pb$_{10}$(PO$_4$)$_6$ is calculated to be stable and has a very small decomposition energy of 3 meV/atom into its competing phases PbO and P$_2$Pb$_3$O$_8$. 
It can be seen that the convex hull  has regions that are very ``flat" in this composition space along the PbO-–P$_{2}$O$_{5}$ isoplethal section. Along these ``flat" regions of the convex hull, many stable or nearly stable ($<$5 meV/atom) compounds exist with near zero decomposition energies. The compounds lying on or near the PbO--P$_{2}$O$_{5}$ convex hull have some structural similarities, but do not follow the kind of formulaic pattern characteristic of a homologous series. The structures all contain PO$_{4}$ tetrahedra, the connectivity of which is dependent on whether there is excess or deficient P. The excess-P end (toweard P$_{2}$O$_{5}$) consists of compounds with a 3D network of fully connected corner-sharing PO$_{4}$ tetrahedra (See Figure~\ref{figs1}). The excess-P compounds PbP$_{2}$O$_{6}$, Pb$_{3}$P$_{4}$O$_{13}$, and Pb$_{2}$P$_{2}$O$_{7}$ have corner-sharing PO$_{4}$ tetrahedra, with connectivity dependent on the P concentration relative to O. For example, the most P-rich ternary compound, PbP$_{4}$O$_{11}$, consists of 2D layers of fully connected corner-sharing PO$_{4}$ tetrahedra, while PbP$_{2}$O$_{6}$ has 1D channels, Pb$_{3}$P$_{4}$O$_{13}$ has separate units of 3 corner-sharing tetrahedra, and Pb$_{2}$P$_{2}$O$_{7}$ has units of 2. On the other hand, Pb$_{3}$P$_{2}$O$_{8}$ and deficient-P compounds Pb$_{10}$P$_{6}$O$_{25}$, Pb$_{4}$P$_{2}$O$_{9}$, and Pb$_{8}$P$_{2}$O$_{13}$ have separated (non-corner-sharing) PO$_{4}$ tetrahedra (See Figure~\ref{figs2}). At the Pb-rich end, PbO consists of 2D layers of edge-sharing square pyramids with Pb atoms lying at the apex and O-free layers. Following this style of coordination, the Pb--P--O ternary compounds consist of Pb-O polyhedra that tend towards square pyramid at the Pb-rich end and octahedra at the Pb-poor end (as required by the excess-P packing of PO$_{4}$ tetrahedra).

Some of the stable compounds along the P$_{2}$O$_{5}$--PbO intersection are isostructural with other compositions in ICSD (Inorganic Crystal Structure Database)~\cite{icsd1}; these are listed in Table~\ref{tab:1}. Correspondingly, some ternary phase spaces have convex hulls that are similar to the Pb--P--O convex hulls, most notably Ca--P--O (Figure~\ref{fig2}(b)) and to a smaller degree Sr--P--O and Ba--P--O. We found two compositions Ca$_4$P$_{6}$O$_{19}$ and CaP$_{4}$O$_{11}$ are stable in the Ca--P--O phase space while their Pb counterparts were not reported in the literature and thus not included in OQMD. We created two new hypothetical Pb$_{4}$P$_{6}$O$_{19}$ and PbP$_{4}$O$_{11}$ compounds through elemental substitution and included them into our Pb--P--O phase diagram. The updated PbO--P$_{2}$O$_{5}$ convex hull is shown in Figure~\ref{fig2}(c) and resembles the CaO–P$_{2}$O$_{5}$ counterpart shown in Figure~\ref{fig2}(d). Through the phase diagram comparison, we discovered a new stable compound PbP$_{4}$O$_{11}$ which is likely to be synthesizable and merits further investigation, but this is out of the scope of this work.

\subsection{2. Pb$_{9}$Cu(PO$_{4}$)$_{6}$O}

We next discuss the Cu-substituted lead phosphate apatite. As shown in Figure~\ref{fig1}(a), there are two different Pb sites in the unit cell. Though the reported Cu substitution sites are Pb(1) sites, we investigated the substitution effects on both Pb(1) (Wyckoff position 4f) and Pb(2) (Wyckoff position 6h) sites. Four unique configurations (structures \#1, \#2, \#3 and \#4) were created when substitution on Pb(1) sites, all of which retain the trigonal P3 space group. However, when substituting on Pb(2) sites, two unique configurations (structures \#5 and \#8) with triclinic P1 space group were created. This was explained by Griffin~\cite{Grifin2023}, where Cu interrupts the hexagonal network when substituted on the Pb(2) sites, which causes a structural rearrangement to the lower P1 symmetry.
We then investigated the differences in formation energies and electronic structures among these six different configurations. The total energies (relative to the lowest energy configuration) for the compounds Pb$_{9}$Cu(PO$_{4}$)$_{6}$O with Cu on Pb(1) sites are -6.484 (10), -6.481 (13), -6.481 (13) and -6.487 (7) eV/atom for structures \#1, \#2, \#3 and \#4 while those on Pb(2) sites are -6.494 (0) and -6.492 (2) eV/atom for structures \#5 and \#8, respectively. This indicates that the Cu substitution on Pb(2) sites is more energetically favorable than on Pb(1) sites by 7 meV/atom or 0.287 eV per formula unit. The ratio of Cu-substitution between Pb(1) and Pb(2) sites at the reported synthesis temperature 1198 K~\cite{Lai2023,Sukbaelee} can be estimated using the sum of Boltzmann formula: $\Sigma$exp(-$\Delta$H$_{Pb(1)}$/kT)/$\Sigma$exp(-$\Delta$H$_{Pb(2)}$/kT) = 0.02. This quick calculation indicates that in equilibrium at synthesis temperatures, Cu-substitution on Pb(1) sites has a much smaller occupancy compared to Pb(2). Despite this, it is important to note that the 0.287 eV energy difference observed here is three times smaller than what is seen in the Pb$_{10-x}$Cu$_{x}$(PO$_{4}$)$_{6}$(OH)$_{2}$ discussed later.

\begin{figure*}[hpt!]
\includegraphics[width=0.7\textwidth]{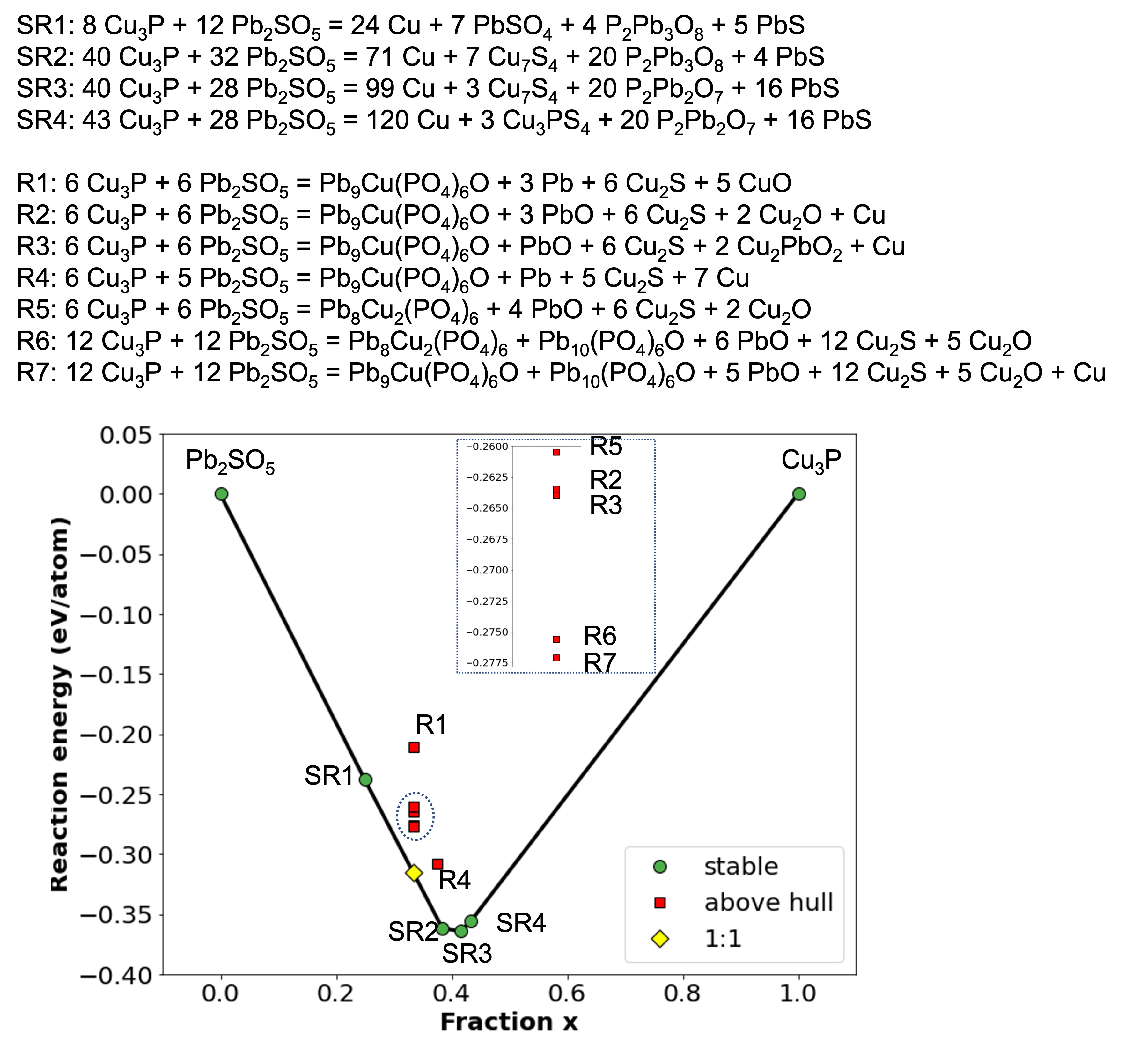}
\caption{Convex hull of the reactions with Cu$_{3}$P and Pb$_{2}$SO$_{5}$ as the reactants. Green circles and red squares indicate that energies of reaction product sare on the hull (stable) and above the convex hull, respectively. The inset plot is a zoom-in area of the five red squares between R5 and R7. The yellow diamond indicates the stable reaction products of 1:1 reactions with Cu$_3$P and Pb$_2$SO$_5$.
} 
\label{fig3}
\end{figure*}

The convex hull distance of Pb$_{9}$Cu(PO$_{4}$)$_{6}$O was calculated as 39 meV/atom above the hull in the lowest energy configuration, competing with ordered ground states CuO and Pb$_{3}$P$_{2}$O$_{8}$. If we consider the ideal configurational entropy from the Pb--Cu mixing on the Pb sublattice and O-Vacancy mixing on the partially occupied oxygen sites, the solid solution of Pb$_{9}$Cu(PO$_{4}$)$_{6}$O does not appear stable enough to compete with the ground states at this fixed composition. The stability energetics calculated here suggests a limited solubility of Cu in the Pb-apatite phase in the equilibrium phase diagram. A special environment with well-tuned chemical potentials of Pb and Cu might be necessary to introduce more Cu-substitution in Pb-apatite.
Oxygen vacancy formation energies were also calculated in both Pb$_{10}$(PO$_{4}$)$_{6}$O and Pb$_{9}$Cu(PO$_{4}$)$_{6}$O structures by removing the lone oxygen atom in these structures. It was calculated as 4.51 eV Pb$_{10}$(PO$_{4}$)$_{6}$O structure. In Pb$_{9}$Cu(PO$_{4}$)$_{6}$O, we calculated the oxygen vacancy formation energies of both Cu substitution on the Pb(1) site as 2.51 eV and on the Pb(2) site as 2.78 eV. The oxygen vacancy formation energy is much lower in Cu-substituted structures, showing that Cu doping increases the possibility of the formation of oxygen vacancies. Therefore, a Pb$_{8}$Cu$_{2}$(PO$_{4}$)$_{6}$ structure might also be favored and will be discussed in Section 5.

We next consider the energetics of some synthesis reaction pathways for LK-99. LK-99 was reported to be synthesized with a molar ratio of 1:1 between two reactants Cu$_{3}$P and Pb$_{2}$SO$_{5}$. The reported synthesis pathways were summarized and discussed in Table~\ref{tab:s1} in the Supplemental Information. Here, we consider four possible reaction equations for the synthesis of Pb$_{9}$Cu(PO$_{4}$)$_{6}$O with 1:1 ratio reactants, namely:

\begin{multline*}
  \textbf{\text{R1: }}  6 \text{ Pb}_2\text{SO}_5 + 6 \text{ Cu}_3\text{P} \rightarrow \\
  \text{ Pb}_9\text{Cu}(\text{PO}_4)_6\text{O} + 3 \text{ Pb} + 6 \text{ Cu}_2\text{S} + 5 \text{ CuO}
\end{multline*}
\begin{multline*}
  \textbf{\text{R2: }}  6 \text{ Pb}_2\text{SO}_5 + 6 \text{ Cu}_3\text{P} \rightarrow \\
  \text{ Pb}_9\text{Cu}(\text{PO}_4)_6\text{O} + 3 \text{ PbO} + 6 \text{ Cu}_2\text{S} + 2 \text{ Cu}_2\text{O} + \text{ Cu}
\end{multline*}
\begin{multline*}
  \textbf{\text{R3: }}  6 \text{ Pb}_2\text{SO}_5 + 6 \text{ Cu}_3\text{P} \rightarrow \\
  \text{ Pb}_9\text{Cu}(\text{PO}_4)_6\text{O} + \text{ PbO} + 6 \text{ Cu}_2\text{S} + 2 \text{ Cu}_2\text{PbO}_2 + \text{ Cu}
\end{multline*}
\begin{multline*}
  \textbf{\text{R7: }}  12 \text{ Pb}_2\text{SO}_5 + 12 \text{ Cu}_3\text{P} \rightarrow \\
  \text{ Pb}_9\text{Cu}(\text{PO}_4)_6\text{O} + \text{ Pb}_{10}\text{P}_6\text{O}_{25} + 5 \text{PbO} + 12 \text{ Cu}_2\text{S} + 5 \text{ Cu}_2\text{O} + \text{ Cu}
  \end{multline*}

\noindent{We also considered one non-1:1 ratio reaction:}
\begin{multline*}
  \textbf{\text{R4: }}  5 \text{ Pb}_2\text{SO}_5 + 6 \text{ Cu}_3\text{P} \rightarrow \\
  \text{ Pb}_9\text{Cu}(\text{PO}_4)_6\text{O} + \text{ Pb} + 5 \text{ Cu}_2\text{S} + 7\text{ Cu}
\end{multline*}

The reaction convex hull is constructed in Figure 3, where we performed the Grand Canonical Linear Programming (GCLP) to minimize free energies along various composition ratios~\cite{Akbar2007,Kirk2012}. The green circles in Figure \ref{fig3} are the stable products of reactions that define the convex hull. The yellow diamond indicates the stable product of one reaction with 1:1 Pb$_{2}$SO$_5$ and Cu$_3$P, which is:

8 Pb$_2$SO$_5$ + 8 Cu$_3$P $\rightarrow$ 17 Cu + Cu$_7$S$_4$ +  2 PbSO$_4$ + 4 P$_2$Pb$_3$O$_8$  + 2 PbS.

All of these five possible reactions have negative reaction energies, which is a necessary condition for a reaction to happen.
The reaction energies are -0.211, -0.263, -0.264, -0.308, and -0.277 eV/atom for R1, R2, R3, R4, and R7, respectively.
When considering 1:1 ratio reactants, the reaction energies follow the trend R1 $>$ R2 $>$ R3 $>$ R7, meaning that R7 is the reaction with the strongest energetic driving force among these four. The product of R7 is 38 meV/atom above the hull, while the product of R4 is 45 meV/atom above the hull, which means 1:1 ratio reactants are preferred when synthesizing this material. The electronic band structures of these six different Pb$_{9}$Cu(PO$_{4}$)$_{6}$O configurations were also calculated and presented in Figure~\ref{figs4} in Supplemental Information. The first four structures, where Cu substitutes Pb(1), uniformly show two half-filled low-energy flat bands at the Fermi level. Among them, the \#1 structure is consistent with previous DFT studies on LK-99 electronic structure~\cite{Cabez23,Junwen2023,Liang23,Rafal2023}  with a comparable band dispersion of $\sim$~0.25 eV. The \#4 structure shows a similar band structure to that of the \#1, featuring an analogous Cu $d^9$ orbital configuration, albeit populated by spin-up electrons. The \#2 and \#3 structures also show this pair-wise parity, with two less spin-split states presenting a band dispersion of only $\sim$~0.09 eV. The much flatter bands in the \#2 and \#3 structures were not noted in the previous DFT studies. Similar band structures between pairs \#1,4 and \#2,3 indicate an equivalent or similar characteristic of the two layers of Pb coordination and bonding environment. On the other hand, according to Kurleto {\it{et\,al.}}~\cite{Rafal2023}, the metallic Pb$_{9}$Cu(PO$_{4}$)$_{6}$O tends to open up the band gap by further lowering the structural symmetry if provided with the opportunity. This is confirmed by the two more energetically favored \#5 and \#8 structures with Pb(2) substitution, which further degenerate to P1 symmetry, showing a large band gap. It is worth noting that in all six possible Pb$_{9}$Cu(PO$_{4}$)$_{6}$O configurations, the half-filled flat bands are always dominated by the minimal $p$-$d$ hybridization between Cu and O, which underscores that the peculiar band characteristics originate from the weak Cu-O bonding enabled by the lattice distortion of Pb$_{10}$(PO$_{4}$)$_{6}$O host, rather than the Pb 6$s^2$ lone-pair states accentuated by Cu-induced lattice distortion.

\begin{figure*}[htp!]
\includegraphics[width=1\textwidth]{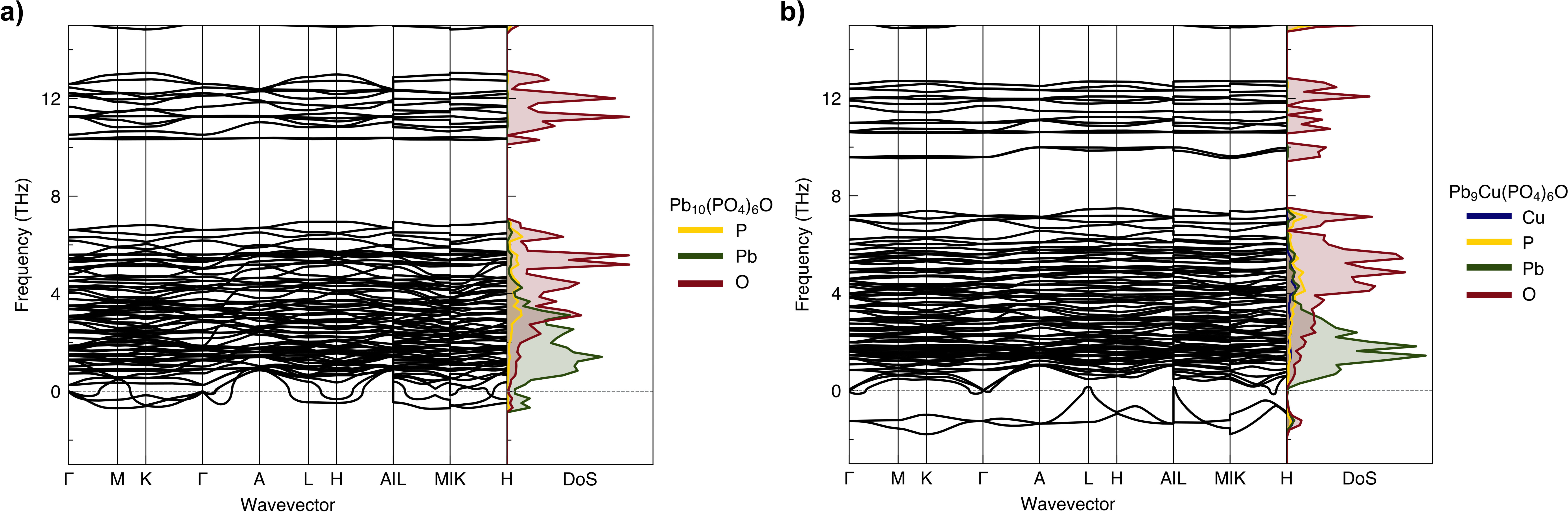}
\caption{Harmonic phonon dispersions and atom-projected density of states of a) Pb$_{10}$(PO$_{4}$)$_{6}$O and b) Pb$_{9}$Cu(PO$_{4}$)$_{6}$O where Cu is substituted on Pb(1) site (\#4 structure).}
\label{fig4}
\end{figure*}

In addition to thermodynamic stability, we also explored the dynamic stability of these structures. We calculated the harmonic phonon dispersions of Pb$_{10}$(PO$_{4}$)$_{6}$O and Pb$_{9}$Cu(PO$_{4}$)$_{6}$O, as shown in Figure~\ref{fig4}(a,b), respectively. Imaginary phonon modes (pictured as negative frequencies in Figure~~\ref{fig4}) show that both structures are not stable to perturbation at T=0K. As illustrated in the atom-projected density of states, the unstable phonon modes in Pb$_{9}$Cu(PO$_{4}$)$_{6}$O mostly arise from the oxygen, Pb, and dopant Cu atoms in the structure, which was also reported in a recent study by Jiang {\it{et\,al.}}~\cite{Jiang23}. Further inclusion of temperature-dependent phonon renormalization might lead to stable phonon modes at finite temperature~\cite{Yi2020}, but this would imply the presence of a phase transition at low temperature – suggesting a more stable structure which is currently unknown.

\subsection{3. Pb$_{10}$(PO$_{4}$)$_{6}$(OH)$_{2}$}
Though LK-99 was reported to be synthesized in a vacuum and did not contain hydrogen in the composition, we investigated the thermodynamic stability of compound Pb$_{10}$(PO$_{4}$)$_{6}$(OH)$_{2}$, which could be synthesized in atmosphere containing hydrogen. Through a structural search, we found at least two compounds Ca$_{10}$(PO$_{4}$)$_{6}$(OH)$_{2}$ and Pb$_{10}$(VO$_{4}$)$_{6}$(OH)$_{2}$ that occupy a similar structure as Pb$_{10}$(PO$_{4}$)$_{6}$O here the difference comes from the inclusion of hydrogen near the oxygen sites along the six-fold axis. In Pb$_{10}$(PO$_{4}$)$_{6}$O, four O sites are 1/4 partially occupied, while in these two compounds, two of these four O sites with the furthest distance between each other are fully occupied by the OH. We performed an elemental substitution and the structure of Pb$_{10}$(PO$_{4}$)$_{6}$(OH)$_{2}$ is shown in Figure~\ref{fig1}(b). The calculated electronic structure of  Pb$_{10}$(PO$_{4}$)$_{6}$(OH)$_{2}$ is shown in Figure~\ref{fig1}(d) with a wide band gap of 3.56 eV. In contrast to Pb$_{10}$(PO$_{4}$)$_{6}$O, Pb$_{10}$(PO$_{4}$)$_{6}$(OH)$_{2}$ shows no isolated valence band near the Fermi level, with only a flat valence band edge consisting of Pb(2) 6$s^2$ lone-pair states and O 2$p$ states. The thermodynamic stability of Pb$_{10}$(PO$_{4}$)$_{6}$(OH)$_{2}$ is evaluated and the phase diagram of Pb--P--O--H is shown in Figure~\ref{figs5}(a) in the Supplemental Information. Pb$_{10}$(PO$_{4}$)$_{6}$(OH)$_{2}$ is calculated as thermodynamically stable (i.e., on the T=0K convex hull) and has a small decomposition energy of 8 meV/atom into its competing phases H$_2$O and Pb$_{10}$(PO$_{4}$)$_{6}$O. This indicates a possibility of the inclusion of hydrogen into LK-99 when exposed to the ambient atmosphere. The isopleth convex hull of P$_{2}$O$_{5}$--H$_2$O--PbO is shown in Figure~\ref{figs5}(b) to visualize the competing phases of Pb$_{10}$(PO$_{4}$)$_{6}$(OH)$_{2}$ in the Pb--P--O--H quaternary phase space.

\subsection{4. Pb$_9$Cu(PO$_4$)$_6$(OH)$_2$}

In this section, we discuss the Cu-substituted hydrogen-containing lead phosphate apatite. Similarly, as the case in section 2, we considered all inequivalent configurations when Cu substituted on Pb(1) or Pb(2) sites. This results in two unique trigonal P3 configurations when substituting on Pb(1) sites and one unique triclinic P1 configuration when substituting on Pb(2) sites. The total energies for the compounds Pb$_9$Cu(PO$_4$)$_6$(OH)$_2$ with Cu on Pb(1) sites are -6.388 and -6.385 eV/atom while those on Pb(2) sites are -6.409 eV/atom. This indicates that the Cu substitution on Pb(2) sites is more energetically favorable than on Pb(1) sites by 21 meV/atom, or 0.86 eV per formula unit, which is close to the previously reported 1.08 eV by Griffin~\cite{Grifin2023}. This large energy difference indicates that Cu is very unlikely to substitute the Pb(1) sites in Pb$_9$Cu(PO$_4$)$_6$(OH)$_2$, at least under equilibrium conditions. The convex hull distance was calculated as 16 meV/atom above the hull for the lowest energy configuration of Pb$_9$Cu(PO$_4$)$_6$(OH)$_2$, competing with H$_2$O, CuO and Pb$_{3}$P$_{2}$O$_{8}$. The convex hull distances of both hydrogen-containing and non-hydrogen-containing compounds with Cu substituted on either Pb(1) or Pb(2) sites are summarized in Table~\ref{tab:2} for comparison. The entropic contribution from limited mixing on Pb(2) sites at 1198K is calculated as 7 meV/atom assuming ideal entropy,  still not sufficient to counter the E$_{\mathrm{hull}}$ value of 16 meV/atom at the synthesis temperature. 
 However, the convex hull energy difference between Pb$_{9}$Cu(PO$_{4}$)$_{6}$O and Pb$_9$Cu(PO$_4$)$_6$(OH)$_2$ (favoring the hydrogen-containing material) again indicates the potential presence of hydrogen in LK-99, which could also enhance the Cu solubility in the Pb-apatite.
 
\begin{table}[htb]
    \resizebox{\columnwidth}{!}{%
        \begin{tabular}{|c|c|c|c|}
            \hline
            Parent Compound & E$_{\mathrm{hull}}$ Cu at Pb(1) & E$_{\mathrm{hull}}$ Cu at Pb(2) \\
            \hline
            Pb$_{10}$(PO$_{4}$)$_{6}$O & 45 meV/atom & 39 meV/atom \\
            \hline
            Pb$_{10}$(PO$_{4}$)$_{6}$(OH)$_{2}$ & 38 meV/atom & 16 meV/atom \\
            \hline
        \end{tabular}
    }
    \caption{The convex hull distances of both hydrogen-containing and not hydrogen-containing compounds with Cu substituted on either Pb(1) or Pb(2) sites.}
    \label{tab:2}
\end{table}

The band structures of these three Pb$_9$Cu(PO$_4$)$_6$(OH)$_2$ configurations were calculated and shown in Figure~\ref{figs6}. The two structures with Cu substitution on Pb(1) site (\#4 and \#5) present Cu-O bonding dominated half-filled flat bands at the Fermi level similar to that of Pb$_{9}$Cu(PO$_{4}$)$_{6}$O. The dispersion of the \#4 and \#5 flat bands are $\sim$0.08 eV and $\sim$0.14 eV, respectively. In the \#1 structure where Cu is substituted on Pb(2) site, the flat band is $\sim$1.6 eV above the Fermi level. Figure~\ref{figs7}(e,f) shows the calculated phonon dispersion of Pb$_{10}$(PO$_{4}$)$_{6}$(OH)$_{2}$ and Pb$_9$Cu(PO$_4$)$_6$(OH)$_2$. They are very similar to the phonon dispersions of Pb$_{10}$(PO$_{4}$)$_{6}$O  and Pb$_{9}$Cu(PO$_{4}$)$_{6}$O in that the Cu doped structures exhibit additional imaginary phonon modes when compared to their parent compounds.

\subsection{5. Pb$_{8}$Cu$_{2}$(PO$_{4}$)$_{6}$}

\begin{figure*}[htp!]
\includegraphics[width=1\textwidth]{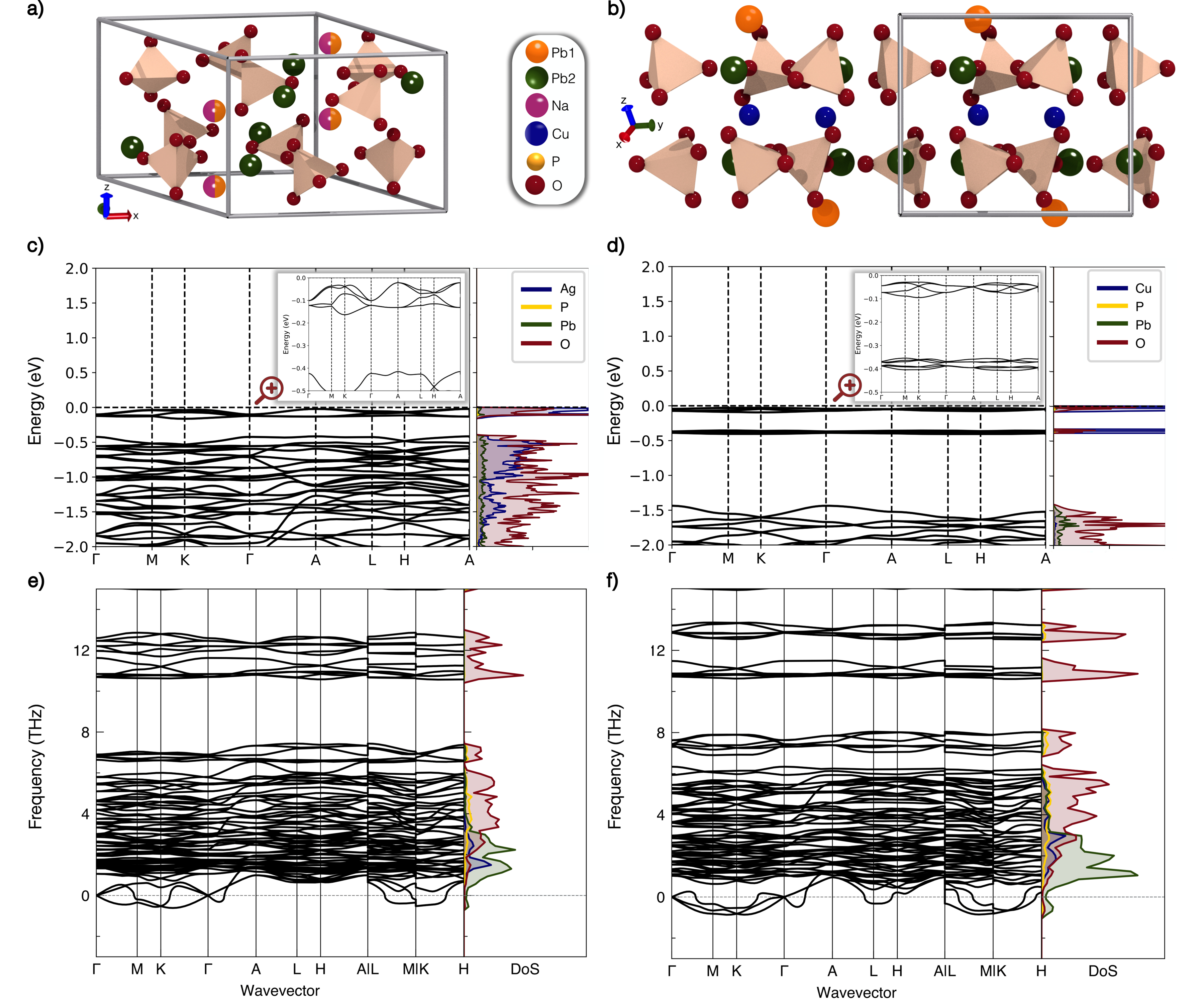}
\caption{Crystal structure of a) Pb$_{8}$Na$_{2}$(PO$5_{4}$)$_{6}$ where two Na and two Pb are mixing at Pb(1) sites and b) the lowest energy configuration of Pb$_{8}$Cu$_{2}$(PO$_{4}$)$_{6}$ where a Cu layer is existing between two layers of PO$_{4}$ tetrahedra. Electronic band structure and density of states of c) Pb$_{8}$Ag$_{2}$(PO$_{4}$)$_{6}$ and d) Pb$_{8}$Cu$_{2}$(PO$_{4}$)$_{6}$ Phonon band structure and atom-projected density of states of e) Pb$_{8}$Ag$_{2}$(PO$_{4}$)$_{6}$ and f) Pb$_{8}$Cu$_{2}$(PO$_{4}$)$_{6}$.}
\label{fig5}
\end{figure*}

The unstable phonon modes observed in either Pb$_{9}$Cu(PO$_{4}$)$_{6}$O or Pb$_9$Cu(PO$_4$)$_6$(OH)$_2$ imply a possible phase transition at low temperature or the existence of a different structure that is dynamically stable at the ground state. The atom-projected phonon density of states shows that the lead, extra oxygen atom, and the Cu dopant may contribute to the imaginary phonon modes at low temperature~\cite{Jiang23}. Also, the oxygen vacancy formation energy calculation suggests that the Cu dopant increases the possibility of the formation of oxygen vacancies. Here we considered removing the extra oxygen in the original Pb$_{10}$(PO$_{4}$)$_{6}$O structure, and then substituting two Pb atoms with two monovalent Cu atoms to balance the charge. Consideration of this novel structure was, in part, motivated by the fact that several alkali metal doped lead apatites have been reported in this structure~\cite{Mullica1986,toumi2008}. One example is Pb$_{8}$Na$_{2}$(PO$_{4}$)$_{6}$ (Figure~\ref{fig5}(a)) where four Pb(1) sites are 50-50 partially occupied by Na and Pb atoms. Substituting Na with Cu atoms in this structure and considering all possibilities within a unit cell, we obtained three unique configurations of Pb$_{8}$Cu$_{2}$(PO$_{4}$)$_{6}$. The lowest energy configuration is shown in Figure~\ref{fig5}(b) where Cu atoms form a layer in the structure between two layers of PO$_{4}$ tetrahedra and the distance between two nearest Cu atoms is 5.56 {\AA}. This distance is 6.61 {\AA} and 3.10 {\AA} respectively in the other two configurations and they are 8 meV/atom and 19 meV/atom higher in energy than the presented one.

The convex hull distance of Pb$_{8}$Cu$_{2}$(PO$_{4}$)$_{6}$ is 31 meV/atom above the hull, competing with Cu$_{2}$O + Pb$_{2}$P$_{2}$O$_{7}$  + Pb$_{3}$P$_{2}$O$_{8}$. The synthesis of the Ag counterpart of this structure was reported in literature~\cite{Ternane2000} and we calculated that using the same structure as Cu. Pb$_{8}$Ag$_{2}$(PO$_{4}$)$_{6}$ is 5 meV/atom above the hull, competing with Ag$_3$PO$_{4}$ + Pb$_{3}$P$_{2}$O$_{8}$. The structure with an Ag layer is also the lowest energy configuration analogous to the Cu case. We considered two possible reaction equations for the synthesis of Pb$_{8}$Cu$_{2}$(PO$_{4}$)$_{6}$ with 1:1 ratio reactants:

\begin{multline*}
  \textbf{\text{R5: }} 6 \text{ Pb}_2\text{SO}_5 + 6 \text{ Cu}_3\text{P} \rightarrow \\ \text{ Pb}_8\text{Cu}_2(\text{PO}_4)_6 + 
  4 \text{ PbO} + 6 \text{ Cu}_2\text{S} + 2 \text{ Cu}_2\text{O}
\end{multline*}
\begin{multline*}
  \textbf{\text{R6: }} 12 \text{ Pb}_2\text{SO}_5 + 12 \text{ Cu}_3\text{P} \rightarrow  \text{ Pb}_8\text{Cu}_2(\text{PO}_4)_6 + \\ 
  \text{ Pb}_{10}\text{P}_6\text{O}_{25} + 6 \text{ PbO} + 12 \text{ Cu}_2\text{S} + 5 \text{ Cu}_2\text{O}
\end{multline*}

As shown in Figure~\ref{fig3}, both R5 and R6 are negative in reaction energies, and R6 is 15 meV/atom lower in energy than R5. Additionally, R6 is 39 meV/atom above the convex hull, which is similar in energy with the most favorable reaction R3 for Pb$_{9}$Cu(PO$_{4}$)$_{6}$O, meaning that both Pb$_{8}$Cu$_{2}$(PO$_{4}$)$_{6}$ are likely to be the product when synthesizing under 1:1 ratio of Pb$_{2}$SO$_{5}$ and Cu$_{3}$P. We suggest that experimental efforts should attempt the formation of the newly predicted Pb$_{8}$Cu$_{2}$(PO$_{4}$)$_{6}$.

The electronic band structures were calculated for both Pb$_{8}$Ag$_{2}$(PO$_{4}$)$_{6}$ (Figure~\ref{fig5}(c)) and Pb$_{8}$Cu$_{2}$(PO$_{4}$)$_{6}$ (Figure~\ref{fig5}(d)) in their most stable configuration. The two compounds are spin-degenerate semiconductors with a wide bandgap of 3.11 eV and 2.05 eV, respectively. Based on the DOS results, the flat valence band edge of Pb$_{8}$Ag$_{2}$(PO$_{4}$)$_{6}$ and Pb$_{8}$Cu$_{2}$(PO$_{4}$)$_{6}$  is dominated by Ag--O and Cu--O bonding states, respectively. The phonon band structures of Pb$_{8}$Ag$_{2}$(PO$_{4}$)$_{6}$ and Pb$_{8}$Cu$_{2}$(PO$_{4}$)$_{6}$ are shown in Figure~\ref{fig5}(e,f), and they once again show dynamic instability at T=0K. However, the lowest phonon frequencies of the imaginary phonon modes are higher in frequency compared to those in the Pb$_{9}$Cu(PO$_{4}$)$_{6}$O structure. Also, no imaginary phonon modes are shown along the $\Gamma$, A and H directions in Pb$_{8}$Cu$_{2}$(PO$_{4}$)$_{6}$ while the imaginary phonon modes can be found along these directions in Pb$_{9}$Cu(PO$_{4}$)$_{6}$O. However, there are three imaginary phonon modes in Pb$_{8}$Cu$_{2}$(PO$_{4}$)$_{6}$ compared to two in Pb$_{9}$Cu(PO$_{4}$)$_{6}$O, which makes it difficult to conclude Pb$_{8}$Cu$_{2}$(PO$_{4}$)$_{6}$ is dynamically more stable than Pb$_{9}$Cu(PO$_{4}$)$_{6}$O. Further investigation of the dynamical stability at finite temperature through renormalized phonon calculations should be performed, as the successful reported experimental synthesis of Pb$_{8}$Ag$_{2}$(PO$_{4}$)$_{6}$ implies that this structure should become dynamically stable at finite temperature.

\section{Conclusions}
In this work, we performed DFT calculations to study the thermodynamic stability, dynamic stability, and electronic structure of LK-99 and its related compounds. Both Pb$_{10}$(PO$_{4}$)$_{6}$O and Pb$_{10}$(PO$_{4}$)$_{6}$(OH)$_{2}$ are calculated to be stable (i.e., on the convex hull) while the Cu-substituted compounds Pb$_{9}$Cu(PO$_{4}$)$_{6}$O and Pb$_9$Cu(PO$_4$)$_6$(OH)$_2$ are unstable (i.e., above the convex hull). Many compounds compete with each other along the PbO--P$_{2}$O$_{5}$ convex hull and they have some structural similarities. The Ca--P--O system has a convex hull similar to the Pb--P--O system, and we discovered a new stable compound Pb$_{4}$P$_{6}$O$_{19}$ through the phase diagram comparison, which completes the phase space of Pb--P--O. The phonon dispersions reveal dynamic instability in all of these four structures, and the oxygen vacancy formation energy calculations suggest that Cu dopant increases the possibility of oxygen vacancy formation. Therefore, we studied a new configuration Pb$_{8}$Cu$_{2}$(PO$_{4}$)$_{6}$, where two Pb(1) sites in the original Pb$_{10}$(PO$_{4}$)$_{6}$O structure are substituted with two monovalent Cu atoms and the extra oxygen is removed from the structure. Though Pb$_{8}$Cu$_{2}$(PO$_{4}$)$_{6}$ is still above the convex hull, the phonon dispersion shows slightly less instability than Pb$_{9}$Cu(PO$_{4}$)$_{6}$O. Several different reaction equations were proposed and their preferences were evaluated through reaction convex hull energies. With a 1:1 ratio of reactants Pb$_{2}$SO$_{5}$ and Cu$_{3}$P, our findings indicate that both Pb$_{8}$Cu$_{2}$(PO$_{4}$)$_{6}$ and Pb$_{9}$Cu(PO$_{4}$)$_{6}$O are likely to be synthesized. Our study proposed a new possible structure for LK-99 and more sophisticated studies (e.g., renormalized anharmonic phonon calculations) on these structures are required to explore their dynamic stabilities at finite temperatures.

\paragraph*{Acknowledgements} 
J.S. acknowledges support from the MRSEC program (DMR-1720319) at the Materials Research Center of Northwestern University (overall leadership of the project, DFT convex hull calculations). D.G., T.L, and C.W. acknowledge financial support received from Toyota Research Institute (TRI) through the Accelerated Materials Design and Discovery program (phonon calculations), S.S and Z.L acknowledge the Department of Energy, Office of Science, Basic Energy Sciences under grant DE-SC0014520 (band structure calculations). S.G. acknowledges the U.S. Department of Commerce and National Institute of Standards and Technology as part of the Center for Hierarchical Materials Design (CHiMaD) under award no. 70NANB14H012 (DFT calculations). A.S.C. acknowledges the financial support to National Agency for Research and Development (ANID)/DOCTORADO BECAS CHILE/2018 - 56180024. T.L. also acknowledges the financial support received from Taiwanese Government Fellowship (Cu-substitution discussion). D.K. acknowledges funding from the International Institute for Nanotechnology (IIN) and the Predictive Science and Engineering Design (PSED) Program at Northwestern University. (DFT calculations). All authors acknowledge the computational resources of the Bridges-2 supercomputer at Pittsburgh Supercomputer Center (PSC) under Grant No. DMR160112 (DFT calculations for band and DOS), and the Quest high-performance computing facility at Northwestern University. This research used the computational resources of the National Energy Research Scientific Computing Center (NERSC), a U.S. Department of Energy Office of Science User Facility located at Lawrence Berkeley National Laboratory, operated under Contract No. DE-AC02-05CH11231 using NERSC award BES-ERCAP23792. 
\clearpage

\paragraph{Data Availability}
We have made the input files corresponding to all structures in this paper available at \url{https://github.com/wolverton-research-group/lead-phosphate-apatites}.

\bibliographystyle{hieeetr}
\bibliography{Ref}

\begin{thebibliography}{10}

\bibitem{Holmes2013}
D.~S. Holmes, A.~L. Ripple, and M.~A. Manheimer, ``{Energy-Efficient
  Superconducting Computing—Power Budgets and Requirements},'' {\em IEEE
  Trans. Appl. Supercond.}, vol.~23, pp.~1701610--–1701610, Jun 2013.

\bibitem{Rote}
D.~M. Rote and L.~R. Johnson, ``{Potential Benefits of Superconductivity to
  Transportation in the United States},'' in {\em Advances in
  Superconductivity: Proceedings of the 1st International Symposium on
  Superconductivity (ISS88), August 28--31, 1988, Nagoya}, pp.~65--70,
  Springer, 1989.

\bibitem{Ali2022}
S.~Ali, ``{Superconductors for Medical Applications},'' {\em MRF},
  pp.~211–--229, Nov 2022.

\bibitem{onnes1991}
H.~K. Onnes, ``Further experiments with liquid helium. c. on the change of
  electric resistance of pure metals at very low temperatures etc. iv. the
  resistance of pure mercury at helium temperatures,'' in {\em Through
  Measurement to Knowledge: The Selected Papers of Heike Kamerlingh Onnes
  1853--1926}, pp.~261--263, Springer, 1991.

\bibitem{Einsenstein1954}
J.~Eisenstein, ``Superconducting elements,'' {\em Rev. Mod. Phys.}, vol.~26,
  pp.~277--–291, Jul 1954.

\bibitem{Bednorz1986}
J.~G. Bednorz and K.~A. M\"{u}ller, ``{Possible high Tc superconductivity in
  the Ba--La--Cu--O system},'' {\em Zeitschrift f\"{u}r Physik B Condensed
  Matter}, vol.~64, pp.~189--–193, Jun 1986.

\bibitem{Subramanian1988}
M.~A. Subramanian, J.~C. Calabrese, C.~C. Torardi, J.~Gopalakrishnan, T.~R.
  Askew, R.~B. Flippen, K.~J. Morrissey, U.~Chowdhry, and A.~W. Sleight,
  ``{Crystal structure of the high-temperature superconductor
  TI$_{2}$Ba$_{2}$CaCu$_{2}$O$_{8}$},'' {\em Nature}, vol.~332,
  pp.~420–--422, Mar 1988.

\bibitem{Drozdov2015}
A.~P. Drozdov, M.~I. Eremets, I.~A. Troyan, V.~Ksenofontov, and S.~I. Shylin,
  ``{Conventional Superconductivity at 203 Kelvin at High Pressures in the
  Sulfur Hydride System},'' {\em Nature}, vol.~525, pp.~73--–76, Aug 2015.

\bibitem{Drozdov22019}
A.~P. Drozdov, P.~P. Kong, V.~S. Minkov, S.~P. Besedin, M.~A. Kuzovnikov,
  S.~Mozaffari, L.~Balicas, F.~F. Balakirev, D.~E. Graf, V.~B. Prakapenka,
  E.~Greenberg, D.~A. Knyazev, M.~Tkacz, and M.~I. Eremets,
  ``{Superconductivity at 250 K in Lanthanum Hydride under High Pressures},''
  {\em Nature}, vol.~569, pp.~528–--531, May 2019.

\bibitem{Sun2019}
Y.~Sun, J.~Lv, Y.~Xie, H.~Liu, and Y.~Ma, ``{Route to a Superconducting Phase
  above Room Temperature in Electron-Doped Hydride Compounds under High
  Pressure},'' {\em Phys. Rev. Lett.}, vol.~123, Aug 2019.

\bibitem{Sukbae23}
S.~Lee, J.-H. Kim, and Y.-W. Kwon, ``{The First Room-Temperature
  Ambient-Pressure Superconductor},'' 2023, 2307.12008v1.

\bibitem{Sukbaelee}
S.~Lee, J.~Kim, H.-T. Kim, S.~Im, S.~An, and K.~H. Auh, ``{Superconductor
  Pb$_{10-x}$Cu$_x$(PO$_4$)$_6$O Showing Levitation at Room Temperature and
  Atmospheric Pressure and Mechanism},'' 2023, 2307.12037v2.

\bibitem{Zhao2014}
S.~Zhao, P.~Gong, S.~Luo, L.~Bai, Z.~Lin, C.~Ji, T.~Chen, M.~Hong, and J.~Luo,
  ``{Deep--Ultraviolet Transparent Phosphates
  RbBa$_{2}$(PO\textsubscript{3})$_{5}$ and
  Rb$_{2}$Ba$_{3}$(P$_{2}$O$_{7}$)$_{2}$ Show Nonlinear Optical Activity from
  Condensation of [PO\textsubscript{4}]$^{3-}$ Units},'' {\em J. Am. Chem.
  Soc.}, vol.~136, pp.~8560--8563, Jun 2014.

\bibitem{Dong2016}
X.~Dong, Y.~Shi, M.~Zhang, Z.~Chen, Q.~Jing, Y.~Yang, S.~Pan, Z.~Yang, and
  H.~Li, ``{The Flexibility of P$_2$O$_7$ Dimers in Soft Structures:
  M$_{2}$CdP$_2$O$_7$ (M = Rb, Cs)},'' {\em Eur. J. Inorg. Chem.}, vol.~2016,
  pp.~2704–--2708, Apr 2016.

\bibitem{Grifin2023}
S.~M. Griffin, ``{Origin of Correlated Isolated Flat Bands in
  Copper--Substituted Lead Phosphate Apatite},'' 2023, 2307.16892v2.

\bibitem{Rafal2023}
R.~Kurleto, S.~Lany, D.~Pashov, S.~Acharya, M.~van Schilfgaarde, and D.~S.
  Dessau, ``{Pb--Apatite Framework as a Generator of Novel Flat-Band CuO Based
  Physics, Including Possible Room Temperature Superconductivity},'' 2023,
  2308.00698v1.

\bibitem{Cabez23}
J.~Cabezas-Escares, N.~F. Barrera, C.~Cardenas, and F.~Munoz, ``{Theoretical
  insight on the LK--99 material},'' 2023, 2308.01135v1.

\bibitem{VASP1}
G.~Kresse and J.~Furthm\"uller, ``Efficient iterative schemes for {\it
  ab-initio} total-energy calculations using a plane-wave basis set,'' {\em
  Phys. Rev. B}, vol.~54, pp.~11169--11186, 1996.

\bibitem{VASP2}
G.~Kresse and J.~Furthm\"{u}ller, ``{Efficiency of \textit{ab-initio} Total
  Energy Calculations for Metals and Semiconductors Using a Plane-Wave Basis
  Set},'' {\em Comput. Mater. Sci.}, vol.~6, no.~1, pp.~15--50, 1996.

\bibitem{PBE}
J.~P. Perdew, K.~Burke, and M.~Ernzerhof, ``Generalized gradient approximation
  made simple,'' {\em Phys. Rev. Lett.}, vol.~78, pp.~1396--1399, 1997.

\bibitem{GGAU}
S.~L. Dudarev, G.~A. Botton, S.~Y. Savrasov, C.~J. Humphreys, and A.~P. Sutton,
  ``{Electron-energy-loss spectra and the structural stability of nickel
  oxide:  An LSDA+U study},'' {\em Phys. Rev. B}, vol.~57,
  pp.~1505--–1509, Jan 1998.

\bibitem{OQMD_2}
S.~Kirklin, J.~E. Saal, B.~Meredig, A.~Thompson, J.~W. Doak, M.~Aykol,
  S.~R{\"{u}}hl, and C.~Wolverton, ``{The Open Quantum Materials Database
  (\texttt{OQMD}): Assessing the accuracy of DFT formation energies},'' {\em
  npj Comput. Mater.}, vol.~1, pp.~1501--1509, 2015.

\bibitem{OQMD_1}
J.~E. Saal, S.~Kirklin, M.~Aykol, B.~Meredig, and C.~Wolverton, ``{Materials
  Design and Discovery with High-Throughput Density Functional Theory: The Open
  Quantum Materials Database (\texttt{OQMD})},'' {\em JOM}, vol.~65, no.~11,
  pp.~1501--1509, 2013.

\bibitem{Phonopy}
A.~Togo, F.~Oba, and I.~Tanaka, ``First-principles calculations of the
  ferroelastic transition between rutile-type and ${\text{cacl}}_{2}$-type
  ${\text{sio}}_{2}\ $ at high pressures,'' {\em Phys. Rev. B}, vol.~78,
  p.~134106, Oct 2008.

\bibitem{Jiahong2021}
J.~Shen, V.~I. Hegde, J.~He, Y.~Xia, and C.~Wolverton, ``{High--Throughput
  Computational Discovery of Ternary Mixed--Anion Oxypnictides},'' {\em Chem.
  Mater.}, vol.~33, pp.~9486--–9500, Dec 2021.

\bibitem{Liang23}
L.~Si and K.~Held, ``{Electronic Structure of the Putative Room-Temperature
  Superconductor Pb$_9$Cu(PO$_4$)$_6$O},'' 2023, 2308.00676v2.

\bibitem{Junwen2023}
J.~Lai, J.~Li, P.~Liu, Y.~Sun, and X.-Q. Chen, ``{First--Principles Study on
  the Electronic Structure of Pb$_{10-x}$Cu$_x$(PO$_4$)$_6$O ($x$=0, 1)},''
  2023, 2307.16040v2.

\bibitem{Mullica1986}
D.~Mullica, H.~O. Perkins, D.~A. Grossie, L.~Boatner, and B.~Sales,
  ``{Structure of Dichromate--Type Lead
  Pyrophosphate,Pb$_{2}$P$_{2}$O$_{7}$},'' {\em J. Solid State Chem.}, vol.~62,
  pp.~371--–376, May 1986.

\bibitem{belokoneva2001}
E.~L. Belokoneva, O.~A. Gurbanova, and O.~V. Dimitrova, ``{ChemInform Abstract:
  A New Crystal Structure of Pb$_{3}$(PO$_{4}$)$_{2}$ and Its Comparison with
  Formula Analogues.},'' {\em ChemInform}, vol.~32, Jan 2001.

\bibitem{Jost1964}
K.~H. Jost, ``{Die Struktur des Bleipolyphosphats [Pb(PO$_{3}$)$_{2}$]xund
  allgemeiner \"{U}berblick \"{u}ber Polyphosphatstrukturen},'' {\em Acta
  Crystallogr.}, vol.~17, pp.~1539--–1544, Dec 1964.

\bibitem{Sukbae2023}
S.~Lee, J.-H. Kim, and Y.-W. Kwon, ``{The First Room--Temperature
  Ambient--Pressure Superconductor},'' 2023, 2307.12008v1.

\bibitem{Krivovichev2003}
S.~V. Krivovichev and P.~C. Burns, ``{Crystal Chemistry of Lead Oxide
  Phosphates: Crystal Structures of Pb$_{4}$O(PO$_{4}$)$_{2}$,
  Pb$_{8}$O$_{5}$(PO$_{4}$)$_{2}$ and Pb$_{10}$(PO$_{4}$)$_{6}$O },'' {\em Z.
  fur Krist. - Cryst. Mater.}, vol.~218, pp.~357--–365, May 2003.

\bibitem{Averbuch1987}
M.~T. Averbuch-Pouchot and A.~Durif, ``{Structure of Lead
  Tetrapolyphosphate},'' {\em Acta Crystallogr. C Struct.}, vol.~43,
  pp.~631–--632, Apr 1987.

\bibitem{Sunn2016}
W.~Sun, S.~T. Dacek, S.~P. Ong, G.~Hautier, A.~Jain, W.~D. Richards, A.~C.
  Gamst, K.~A. Persson, and G.~Ceder, ``{The Thermodynamic Scale of Inorganic
  Crystalline Metastability},'' {\em Sci. Adv.}, vol.~2, Nov 2016.

\bibitem{icsd1}
G.~Bergerhoff, R.~Hundt, R.~Sievers, and I.~D. Brown, ``The inorganic crystal
  structure database,'' {\em J. Chem. Inf. Comput. Sci.}, vol.~23, pp.~66--69,
  May 1983.

\bibitem{Lai2023}
J.~Lai, J.~Li, P.~Liu, Y.~Sun, and X.-Q. Chen, ``{First--Principles Study on
  the Electronic Structure of Pb$_{10-x}$Cu$_{x}$(PO$_{4}$)$_{6}$O(x$=$0,
  1)},'' {\em J. Mater. Sci. Technol.}, Aug 2023.

\bibitem{Akbar2007}
A.~R.~Akbarzadeh, V.~Ozoli\c{n}\u{s}, and C.~Wolverton, ``{First--Principles
  Determination of Multicomponent Hydride Phase Diagrams: Application to the
  Li--Mg--N--H System},'' {\em Adv. Mater.}, vol.~19, no.~20, pp.~3233--3239,
  2007.

\bibitem{Kirk2012}
S.~Kirklin, B.~Meredig, and C.~Wolverton, ``{High--Throughput Computational
  Screening of New Li--Ion Battery Anode Materials},'' {\em Adv. Energy.
  Mater.}, vol.~3, no.~2, pp.~252--262, 2013.

\bibitem{Jiang23}
Y.~Jiang, S.~B. Lee, J.~Herzog-Arbeitman, J.~Yu, X.~Feng, H.~Hu,
  D.~C\u{a}lug\u{a}ru, P.~S. Brodale, E.~L. Gormley, M.~G. Vergniory,
  C.~Felser, S.~Blanco-Canosa, C.~H. Hendon, L.~M. Schoop, and B.~A. Bernevig,
  ``{Pb$_9$Cu(PO4)$_6$(OH)$_2$: Phonon bands, Localized Flat Band Magnetism,
  Models, and Chemical Analysis},'' 2023, 2308.05143v1.

\bibitem{Yi2020}
Y.~Xia, V.~I. Hegde, K.~Pal, X.~Hua, D.~Gaines, S.~Patel, J.~He, M.~Aykol, and
  C.~Wolverton, ``{High--Throughput Study of Lattice Thermal Conductivity in
  Binary Rocksalt and Zinc Blende Compounds Including Higher-Order
  Anharmonicity},'' {\em Phys. Rev. X}, vol.~10, Nov 2020.

\bibitem{toumi2008}
M.~TOUMI and T.~MHIRI, ``{Crystal Structure and Spectroscopic Studies of
  Na$_2$Pb$_8$(PO$_4$)$_6$},'' {\em J. Ceram. Soc. Jpn.}, vol.~116, no.~1356,
  pp.~904–--908, 2008.

\bibitem{Ternane2000}
R.~Ternane, M.~Ferid, N.~Kbir-Ariguib, and M.~Trabelsi-Ayedi, ``{The Silver
  Lead Apatite Pb$_{8}$Ag$_{2}$(PO$_{4}$)$_{6}$: Hydrothermal Preparation},''
  {\em J. Alloys Compd.}, vol.~308, pp.~83--–86, Aug 2000.

\bibitem{LiLiu2023}
L.~Liu, Z.~Meng, X.~Wang, H.~Chen, Z.~Duan, X.~Zhou, H.~Yan, P.~Qin, and
  Z.~Liu, ``{Semiconducting transport in Pb$_{10-x}$Cu$_x$(PO$_4$)$_6$O
  sintered from Pb$_2$SO$_5$ and Cu$_3$P},'' 2023, 2307.16802v1.

\bibitem{Kapil2023}
K.~Kumar, N.~K. Karn, and V.~P.~S. Awana, ``{Synthesis of Possible Room
  Temperature Superconductor LK--99:Pb$_9$Cu(PO$_4$)$_6$O},'' 2023,
  2307.16402v1.

\bibitem{Hou23}
Q.~Hou, W.~Wei, X.~Zhou, Y.~Sun, and Z.~Shi, ``{Observation of zero resistance
  above 100$^\circ$ K in Pb$_{10-x}$Cu$_x$(PO$_4$)$_6$O},'' 2023, 2308.01192v1.

\bibitem{Guo23}
K.~Guo, Y.~Li, and S.~Jia, ``{Ferromagnetic Half Levitation of LK--99--Like
  Synthetic Samples},'' 2023, 2308.03110v1.

\bibitem{kkapil2}
K.~Kumar, N.~K. Karn, Y.~Kumar, and V.~P.~S. Awana, ``{Absence of
  Superconductivity in LK--99 at Ambient Conditions},'' 2023, 2308.03544v3.

\bibitem{Evans}
H.~T. Evans, ``{Copper Coordination in Low Chalcocite and Djurleite and Other
  Copper-Rich Sulfides},'' {\em American Mineralogist}, vol.~66, pp.~807--818,
  08 1981.

\bibitem{white1964}
W.~B. White and R.~Roy, ``{Phase Relations in the System Lead-Oxygen},'' {\em
  J. Am. Ceram.}, vol.~47, pp.~242--249, May 1964.

\bibitem{ZZhang2016}
J.~Zhang and H.~W. Richardson, ``{Copper Compounds},'' {\em Ullmann's encycl.
  ind. chem.}, pp.~1–--31, May 2016.

\bibitem{Dreizin2000}
E.~Dreizin, ``{Phase Changes in Metal Combustion},'' {\em PECS}, vol.~26,
  pp.~57–--78, Feb 2000.

\bibitem{cancer2005}
M.~\u{C}an\u{c}arevi\, {c}, M.~Zinkevich, and F.~Aldinger, ``{Enthalpy of
  Formation of Cu$_{2}$PbO$_{2}$ and Revision of the Cu$_{2}$O--PbO Phase
  Diagram},'' {\em MSF}, vol.~494, pp.~67–--72, Sep 2005.

\bibitem{Chongad2016}
L.~S. Chongad, A.~Sharma, M.~Banerjee, and A.~Jain, ``{Synthesis of Lead
  Sulfide Nanoparticles by Chemical Precipitation Method},'' {\em J. Phys.
  Conf. Ser.}, vol.~755, p.~012032, Oct 2016.

\bibitem{Sadov2013}
S.~Sadovnikov and A.~Gusev, ``{Structure and Properties of PbS Films},'' {\em
  J. Alloys Compd.}, vol.~573, pp.~65–--75, Oct 2013.

\bibitem{silverman1966}
M.~S. Silverman, ``{High-Pressure (70-kbar) Synthesis of New Crystalline Lead
  Dichalcogenides},'' {\em Inorg. Chem.}, vol.~5, pp.~2067–--2069, Nov 1966.

\end{thebibliography}

\appendix
\onecolumngrid

\pagebreak
\section*{Supplemental Information}
\renewcommand{\thesubsection}{\Alph{subsection}}
\renewcommand{\thefigure}{S\arabic{figure}} 
\renewcommand{\thetable}{S\arabic{table}} 

\subsection{Review of experimental conditions for LK-99 synthesis:}

\FloatBarrier
\setcounter{table}{0}
\begin{table*}[htpb!]
  \centering
  \tiny
    \begin{tabularx}{\textwidth}{|c|X|X|X|X|X|X|}
        \hline
        \rule{0pt}{10pt}\textbf{Exp.}[Ref.] & Pb$_{2}$SO$_{5}$:Cu$_{3}$P & Balanced & Impurities & Reactants Mixing & Reaction & Observations \\
        & molar ratio & Reaction & in Products & Conditions & Conditions & \\
        \hline
        \rule{0pt}{10pt}\textbf{1}~\cite{Sukbaelee,Sukbae23} & 1:1 (Fig.1) & Not shown & Cu$_{2}$S, Cu(0), Cu(II) by XRD and XPS& Not explained & Vaccum Sealed (10$^{-3}$ Torr), 925 $^\circ$C, 5-20 h & From XPS product ratio is Pb:Cu = 10:0.9.  Not clear if Lanarkite synthesis was exposed to air or not. (cf. main text vs Fig 1 (b)). Cu$_2$S is Cu(I), but not mentioned in XPS analysis. \\
        \hline
        \rule{0pt}{10pt}\textbf{2}~\cite{LiLiu2023} & 1:1 & Not shown & Not mentioned & Not explained (by pictures, it is atmospheric pressure and exposure to air) & Vacuum sealed (10$^{-5}$ Torr), 925 $^\circ$C, 10 h & -- \\
        \hline
        \rule{0pt}{10pt}\textbf{3}~\cite{Kapil2023} & 1:1 & Not shown & Cu$_2$S by XRD & Not explained & Vacuum sealed, 925 $^\circ$C, 10 h & -- \\
        \hline
        \rule{0pt}{10pt}\textbf{4}~\cite{Hou23} & 1:1 & Not shown & Cu$_{2}$S (lower qty.) & -Glovebox filled with Argon & Compressed sample (6GPa), vacuum sealed, 925 $^\circ$C,  & Final product is independent of the use of Pb$_2$SO$_5$ or Pb$_4$SO$_6$ as reactants.\\
        & 4.5:6 & Not shown & Cu$_{2}$S & The air-exposed mixing reactants for lanarkite synthesis generate Pb$_4$SO$_6$ & (ratio 1.5 $^\circ$C/min), 10 h and 20h & Ratio 1:1 is the key for getting similar results to Exp. 1 (Korean group) \\
        \hline
        \rule{0pt}{10pt}\textbf{5}~\cite{Guo23} & Not specified, but refer to Exp. 1  & Not shown & Cu$_2$S: described as red transparent crystals & Not explained & Not detailed. Refer to Exp.1 & -- \\
        \hline
        \rule{0pt}{10pt}\textbf{6}~\cite{kkapil2} & 1:1 & 5Pb$_2$SO$_5$ + 6Cu$_3$P --> Pb$_9$CuP$_6$O$_{25}$ + 5Cu$_2$S + Pb + 7Cu & Cu$_2$S & Not explained & Vacuum sealed, 925 C, 10 h & Molar ratio of the experiment is not the same as the proposed reaction \\
        \hline
        
    \end{tabularx}
    \caption{Summary of Pb$_{2}$SO$_{5}$ and Cu$_{3}$P reaction}
    \label{tab:s1}
\end{table*}
\FloatBarrier

\subsection{Possible reaction products considering experimental conditions for LK-99 synthesis:}
\begin{itemize}
\item Pb$_{10-x}$Cu$_x$P$_6$O$_{25}$, $(0.9 < x < 1.1)$. The range for Cu concentration is not explained in detail in the original LK-99 papers. According to XPS analysis together with relative atomic sensitivity, the product ratio Pb:Cu = 10:0.9, at least for one sample.

\item Cu$_2$S has been observed in many experiments (see Table~\ref{tab:s1}). Although it is 10 meV/atom above the hull in OQMD, it could be stabilized under a high concentration of Cu~\cite{Evans}. Since the ratio of Cu:S is 3:1 within the reactants of the typical LK-99 synthesis, the formation of Chalcocite (Cu$_2$S) + Cu could be favored.

\item PbO is in the convex hull of OQMD. Qiang Hou~ {\it{et\,al.}}~\cite{Hou23} showed that when a ratio 1:1 is used for either Pb$_2$SO$_5$ (stable) or Pb$_3$SO$_6$ (10 meV/atom above the hull) as reactants with Cu$_3$P, the products are the same. Since Pb$_3$SO$_6$ phase competes with Pb$_2$SO$_5$ + PbO phases, it is possible that PbO could be formed. Furthermore, it is known that if PbO is favored at temperatures above 600 $^\circ$C in the presence of air~\cite{white1964} and in the typical LK-99 synthesis under vacuum conditions and with a 1:1 molar ratio for the reactants, there is O-excess that needs to be consumed.

\item Cu$_2$O is in the convex hull of OQMD but is not mentioned in any experimental report. Cu$_2$O can be generated by an equimolar reaction of Cu + CuO heated to 800-900 $^\circ$C in an inert atmosphere and allowed to cool~\cite{ZZhang2016}, just as LK-99 synthesis conditions. Additionally, since all reactions happened under vacuum conditions and the product is cooled down before opening the quartz tubes, no release of S or O should happen; then O must be part of the final impurity compositions. With the low O excess in the reaction environment and other competing oxide phases (like PbO), the most probable impurity phases, based on the Cu-O phase diagram~\cite{Dreizin2000}, are Cu$_2$O + Cu.

\item Cu$_2$PbO$_2$ is on the convex hull of OQMD but is not mentioned in any experimental report. It can be generated by the reaction between PbO + Cu$_2$O, especially when the PbO:Cu$_2$O ratio is higher than 50\%~\cite{cancer2005}.
\end{itemize}

\section{Discarded possible reaction products due to experimental conditions for LK-99 synthesis}

\begin{itemize}

\item PbS is produced with the assistance of H$_2$S gas~\cite{Chongad2016}or as thin films using wet chemistry~\cite{Sadov2013}.
\item PbS$_2$ is generated under high-pressure conditions~\cite{silverman1966}.
\item P compounds. Since the molar ratio of reactants is generally 1:1, P is probably consumed completely during the generation of Pb$_{10-x}$Cu$_x$P$_6$O$_{25}$.
\end{itemize}
\clearpage
\section{Possible balanced chemical reactions equations}

Based on the previous information, the molar ratio between reactants used is 1:1. The following balanced chemical reactions were considered for the analysis in the main text of this work. See the discussion in Sections 2 and 5 and about the implications of them in Figure~\ref{fig3}.
\vspace{5mm}

General balanced reaction with $x=1$:
\begin{equation*}
6 \, \text{Pb}_2 \text{SO}_5 + 6 \, \text{Cu(I)}_3 \text{P} \rightarrow \text{Pb}_9 \text{Cu(II)} \text{P}_6 \text{O}_{25} + 3 \, \text{Pb} + 6 \, \text{S} + 5 \, \text{O} + 17 \, \text{Cu}
\end{equation*}

General balanced reaction with $x=2$:
\begin{equation*}
6 \, \text{Pb}_2 \text{SO}_5 + 6 \, \text{Cu(I)}_3 \text{P} \rightarrow \text{Pb}_8 \text{Cu(I)}_2 \text{P}_6 \text{O}_{24} + 4 \, \text{Pb} + 6 \, \text{S} + 6 \, \text{O} + 16 \, \text{Cu}
\end{equation*}

\vspace{5mm}

Possible balanced equation of reaction with x=1:
\begin{itemize}
    \item \textbf{R1}:
    \begin{equation*}
    6 \, \text{Pb}_2 \text{SO}_5 + 6 \, \text{Cu(I)}_3 \text{P} \rightarrow \text{Pb}_9 \text{Cu(II)} \text{P}_6 \text{O}_{25} + 3 \, \text{Pb} + 6 \, \text{Cu}_2 \text{S} + 5 \, \text{CuO} 
    \end{equation*}

    \item \textbf{R2}:
    \begin{equation*}
    6 \, \text{Pb}_2 \text{SO}_5 + 6 \, \text{Cu(I)}_3 \text{P} \rightarrow \text{Pb}_9 \text{Cu(II)} \text{P}_6 \text{O}_{25} + 3 \, \text{PbO} + 6 \, \text{Cu}_2 \text{S} + 2 \, \text{Cu}_2 \text{O} + \text{Cu}
    \end{equation*}

    \item \textbf{R3}:
    \begin{equation*}
    6 \, \text{Pb}_2 \text{SO}_5 + 6 \, \text{Cu(I)}_3 \text{P} \rightarrow \text{Pb}_9 \text{Cu(II)} \text{P}_6 \text{O}_{25} + \text{PbO} + 6 \, \text{Cu}_2 \text{S} + 2 \, \text{Cu}_2 \text{PbO}_2 + \text{Cu}
    \end{equation*}

    \item \textbf{R4}:
    \begin{equation*}
    5 \, \text{Pb}_2 \text{SO}_5 + 6 \, \text{Cu(I)}_3 \text{P} \rightarrow \text{Pb}_9 \text{Cu(II)} \text{P}_6 \text{O}_{25} + \text{Pb} + 5 \, \text{Cu}_2 \text{S} + 7 \, \text{Cu}
    \end{equation*}
\end{itemize}

\vspace{5mm}

Possible balanced equation of reaction with x=2:
\begin{itemize}
    \item \textbf{R5}:
    \begin{equation*}
    6 \, \text{Pb}_2 \text{SO}_5 + 6 \, \text{Cu(I)}_3 \text{P} \rightarrow \text{Pb}_8 \text{Cu(I)}_2 \text{P}_6 \text{O}_{24} + 4 \, \text{PbO} + 6 \, \text{Cu}_2 \text{S} + 2 \, \text{Cu}_2 \text{O}
    \end{equation*}
\end{itemize}

\vspace{5mm}

Other possibilities:
\begin{itemize}
    \item \textbf{R6}, by considering the formation of some oxygen vacancies ($x=0$ \& $x=2$):
    \begin{equation*}
    12 \, \text{Pb}_2 \text{SO}_5 + 12 \, \text{Cu(I)}_3 \text{P} \rightarrow \text{Pb}_8 \text{Cu(I)}_2 \text{P}_6 \text{O}_{24} + \text{Pb}_{10} \text{P}_6 \text{O}_{25} + 6 \, \text{PbO} + 12 \, \text{Cu}_2 \text{S} + 5 \, \text{Cu}_2 \text{O}
    \end{equation*}

    \item \textbf{R7}, not considering the formation oxygen vacancies ($x=0$ \& $x=1$):
    \begin{equation*}
    12 \, \text{Pb}_2 \text{SO}_5 + 12 \, \text{Cu(I)}_3 \text{P} \rightarrow \text{Pb}_9 \text{Cu(II)} \text{P}_6 \text{O}_{25} + \text{Pb}_{10} \text{P}_6 \text{O}_{25} + 5 \, \text{PbO} + 12 \, \text{Cu}_2 \text{S} + 5 \, \text{Cu}_2 \text{O} + \text{Cu}
    \end{equation*}
\end{itemize}

\setcounter{figure}{0}
\begin{figure*}[ht!]
\includegraphics[width=1\textwidth]{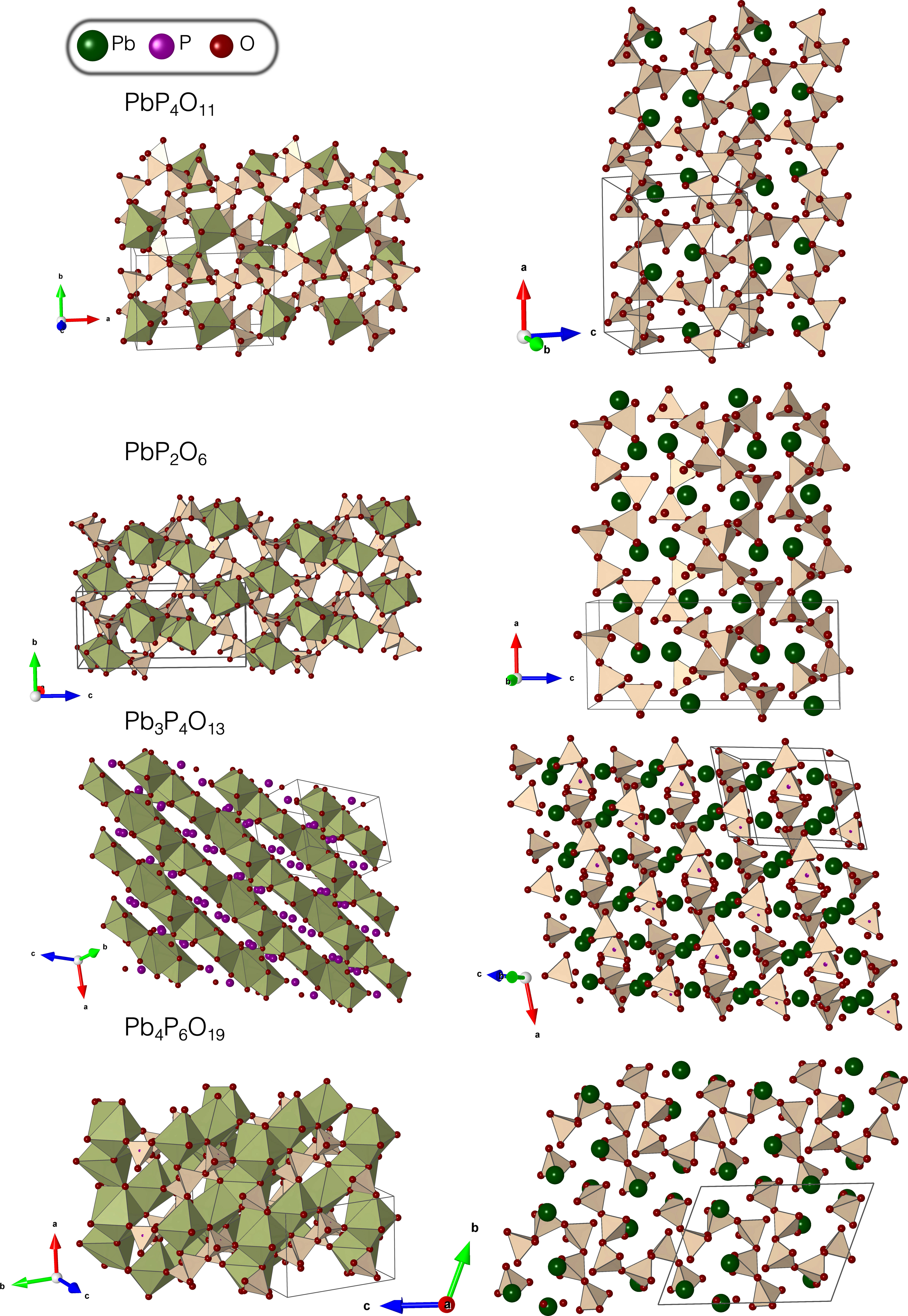}
\caption{Crystal structures of P-rich Compounds.} 
\label{figs1}
\end{figure*}

\begin{figure*}[ht!]
\includegraphics[width=0.9\textwidth]{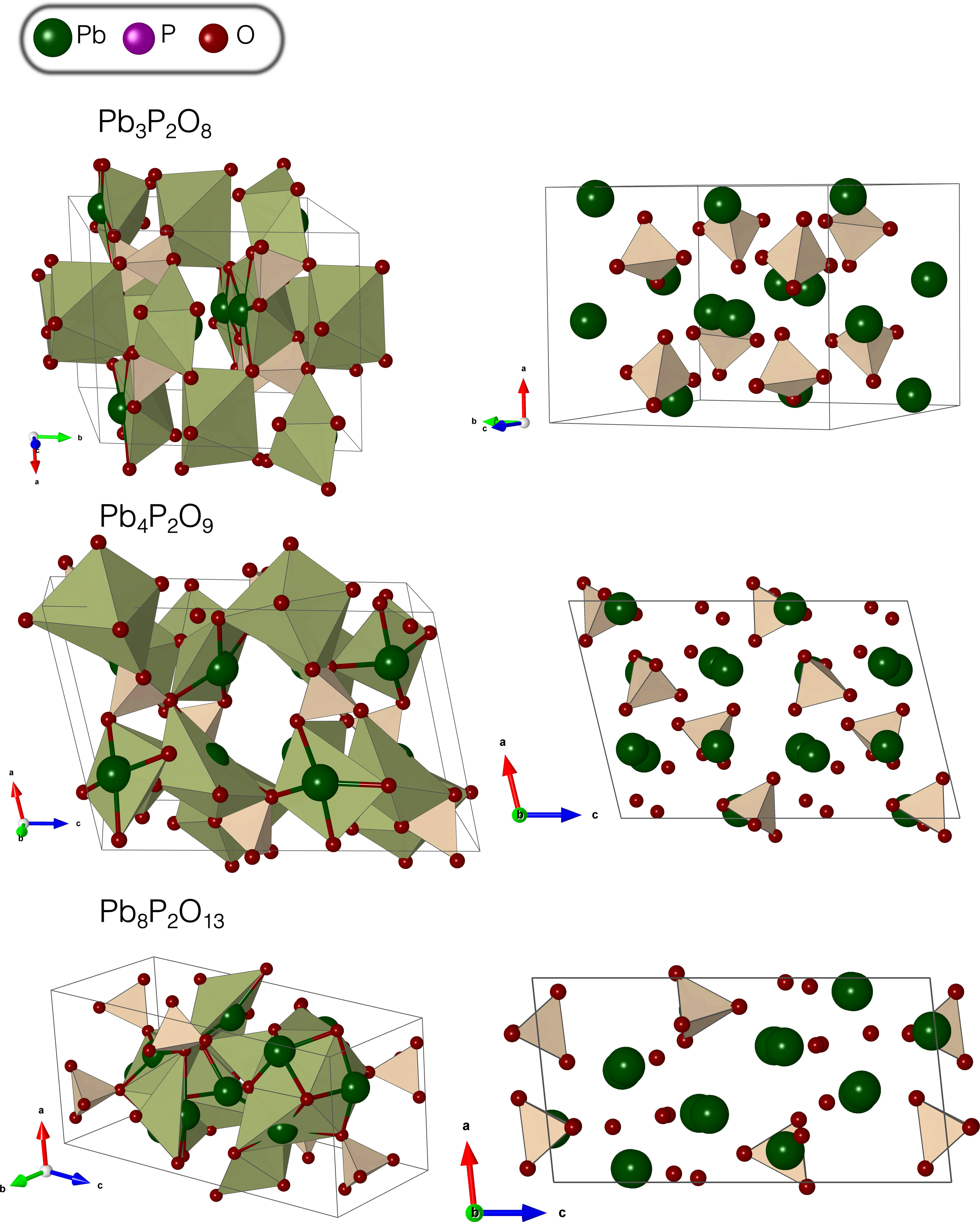}
\caption{Crystal structures of Pb-rich Compounds.} 
\label{figs2}
\end{figure*}

\begin{figure*}[ht!]
\includegraphics[width=0.8\textwidth]{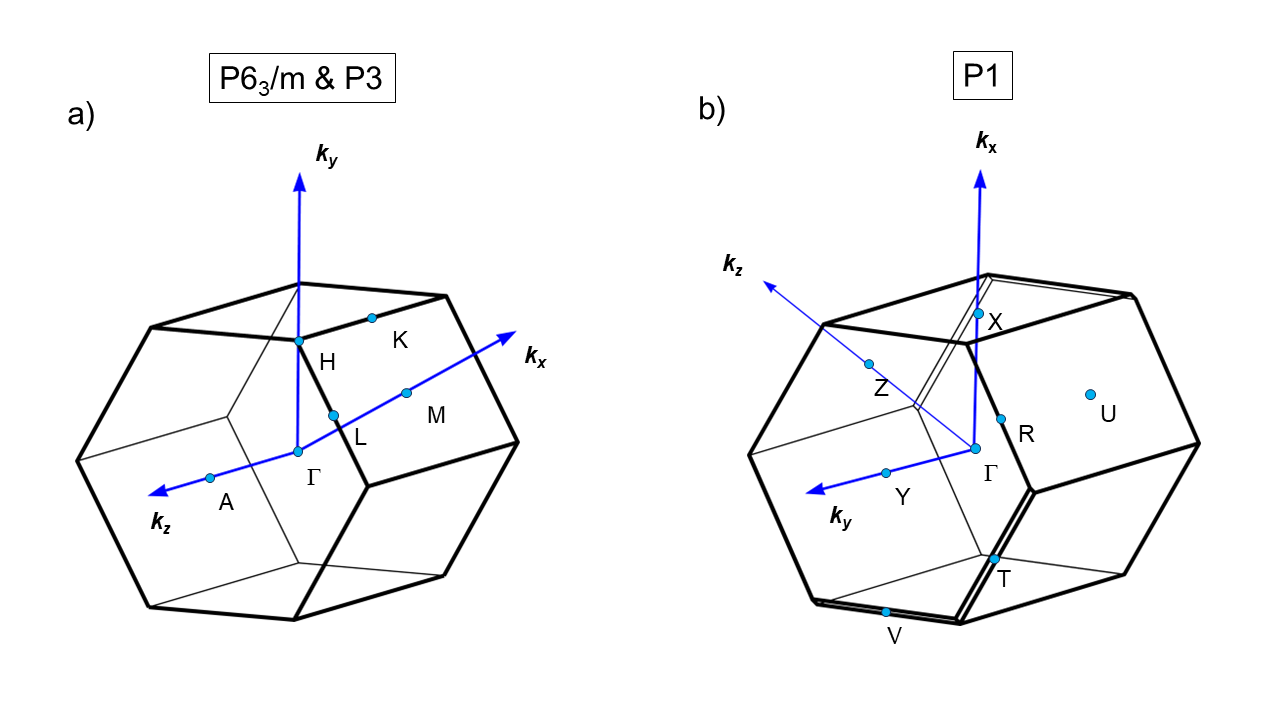}
\caption{First Brillouin zone of P6$_{3}$/m, P3, and P1 space groups.}
\label{figs3}
\end{figure*}

\begin{figure*}[ht!]
\includegraphics[width=1\textwidth]{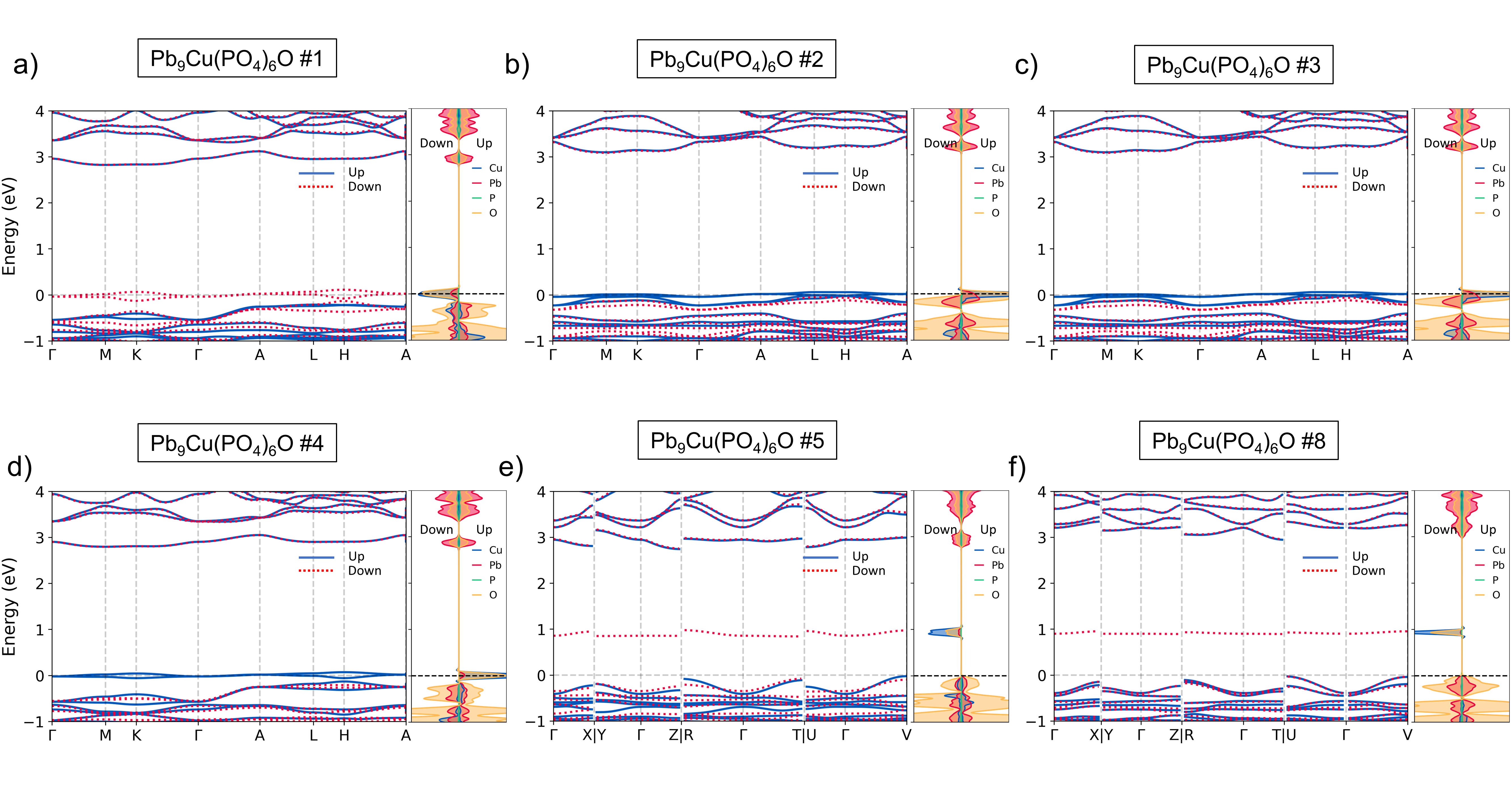}
\caption{Spin-polarized electronic structures and density of states of the six different Pb$_{9}$Cu(PO$_{4}$)$_{6}$O configurations. The total energies are -6.484, -6.481, -6.481, -6.487 eV/atom for \#1, \#2, \#3 and \#4 structures respectively, where Cu is substituted on Pb(1) site. The total energies are -6.494 and -6.492 eV/atom for \#5 and \#8 structures respectively, where Cu is substituted on Pb(2) site.} 
\label{figs4}
\end{figure*}

\begin{figure*}[ht!]
\includegraphics[width=1\textwidth]{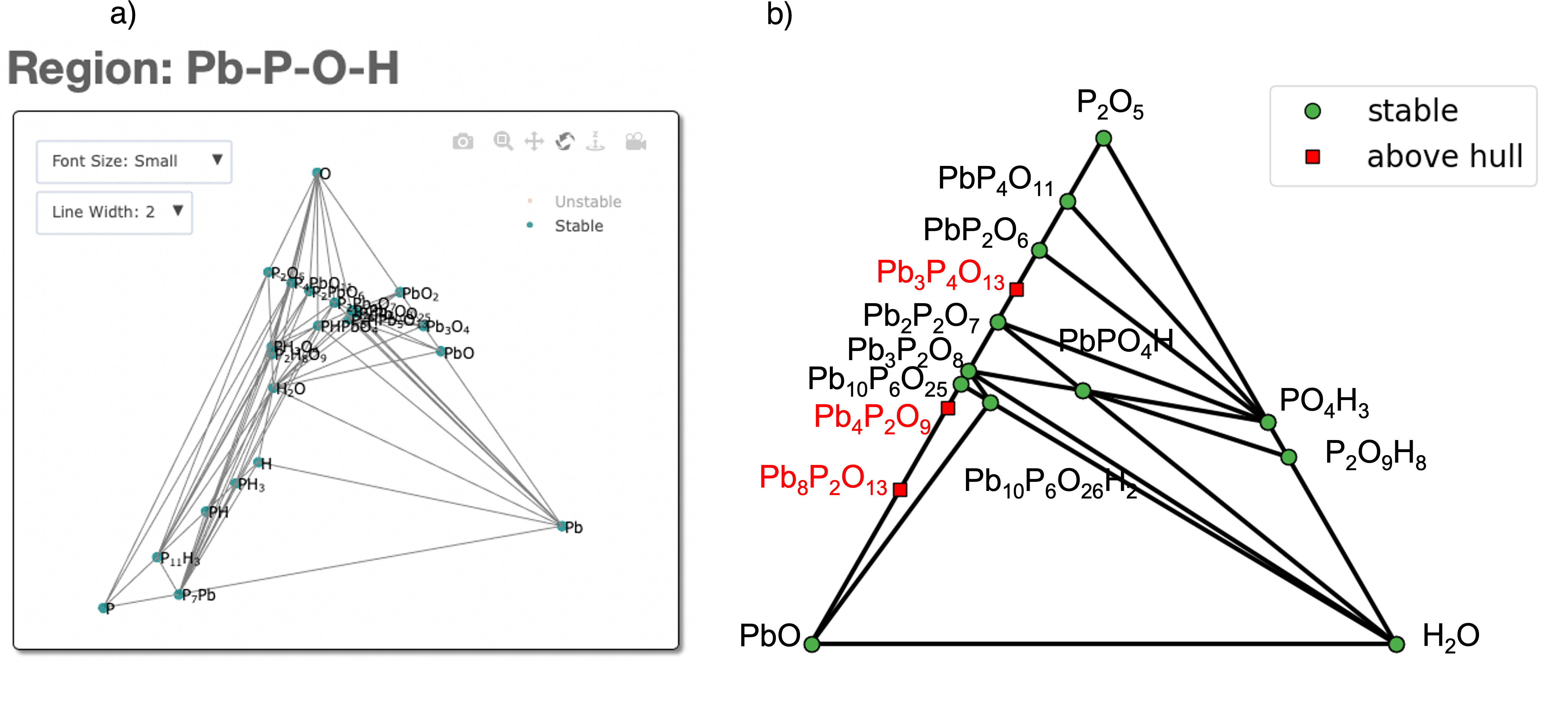}
\caption{The a) Pb--P--O--H phase space and b) convex hull of P$_{2}$O$_{5}$–H$_2$–PbO.}
\label{figs5}
\end{figure*}

\begin{figure*}[ht!]
\includegraphics[width=1\textwidth]{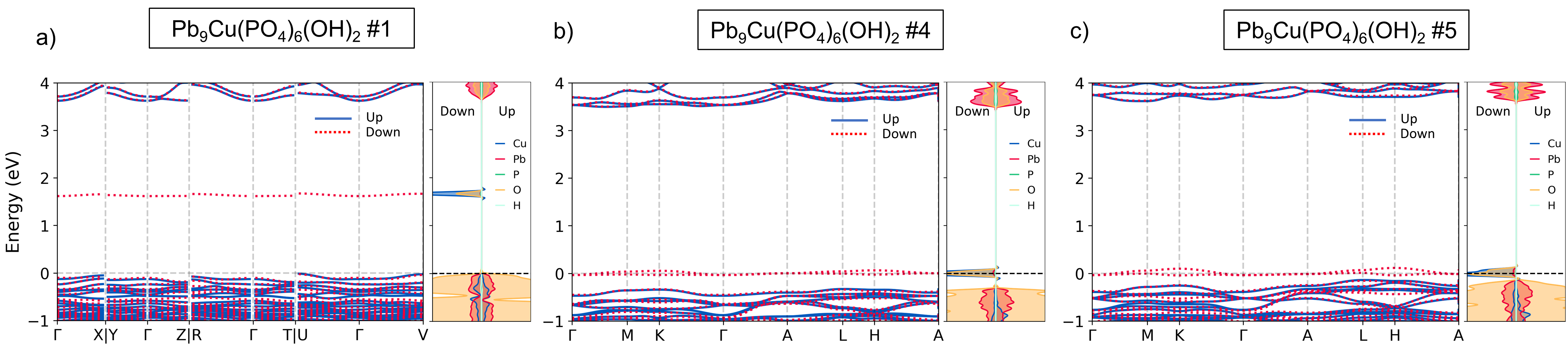}
\caption{Spin-polarized electronic structures and density of states of the three different Pb$_9$Cu(PO$_4$)$_6$(OH)$_2$ configurations. The total energies are -6.388 and -6.385 eV/atom for \#4 and \#5 structures respectively, where Cu is substituted on Pb(1) site. The total energies are -6.409 eV/atom for the \#1 structure, where Cu is substituted on Pb(2) site.}
\label{figs6}
\end{figure*}

\begin{figure*}[ht!]
\includegraphics[width=1\textwidth]{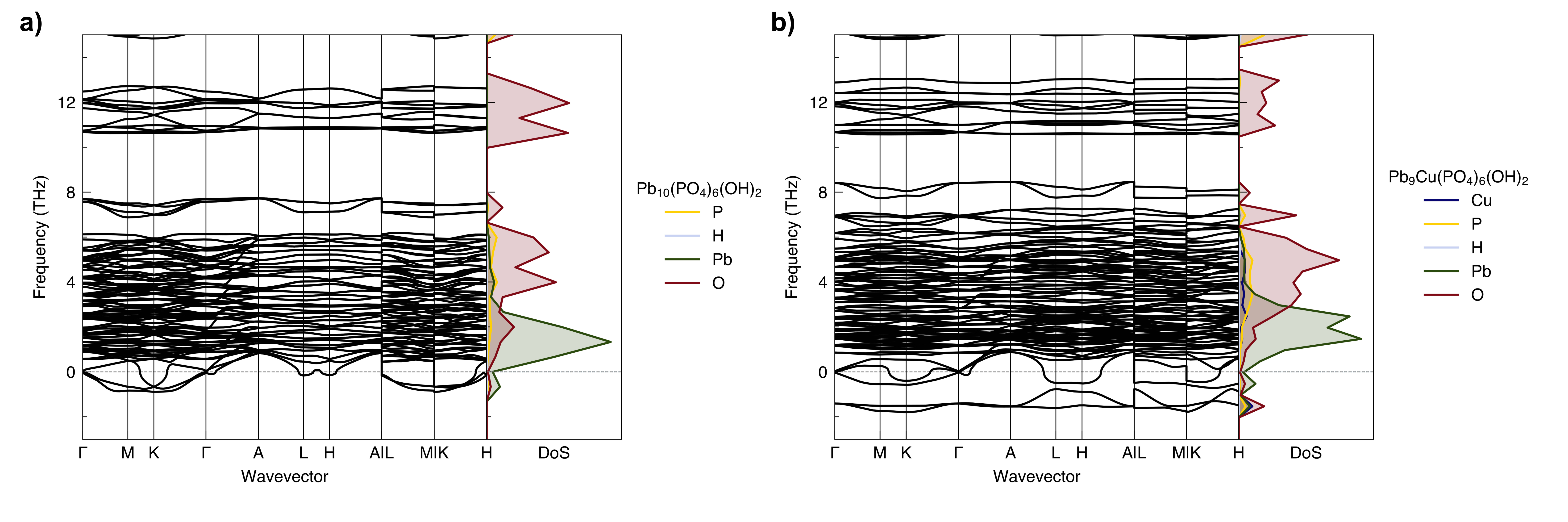}
\caption{Harmonic phonon dispersions of a) Pb$_{10}$(PO$_{4}$)$_{6}$(OH)$_{2}$ and b) Pb$_9$Cu(PO$_4$)$_6$(OH)$_2$ where Cu is substituted on Pb(1) site (\#4 structure).}
\label{figs7}
\includegraphics[width=1\textwidth]{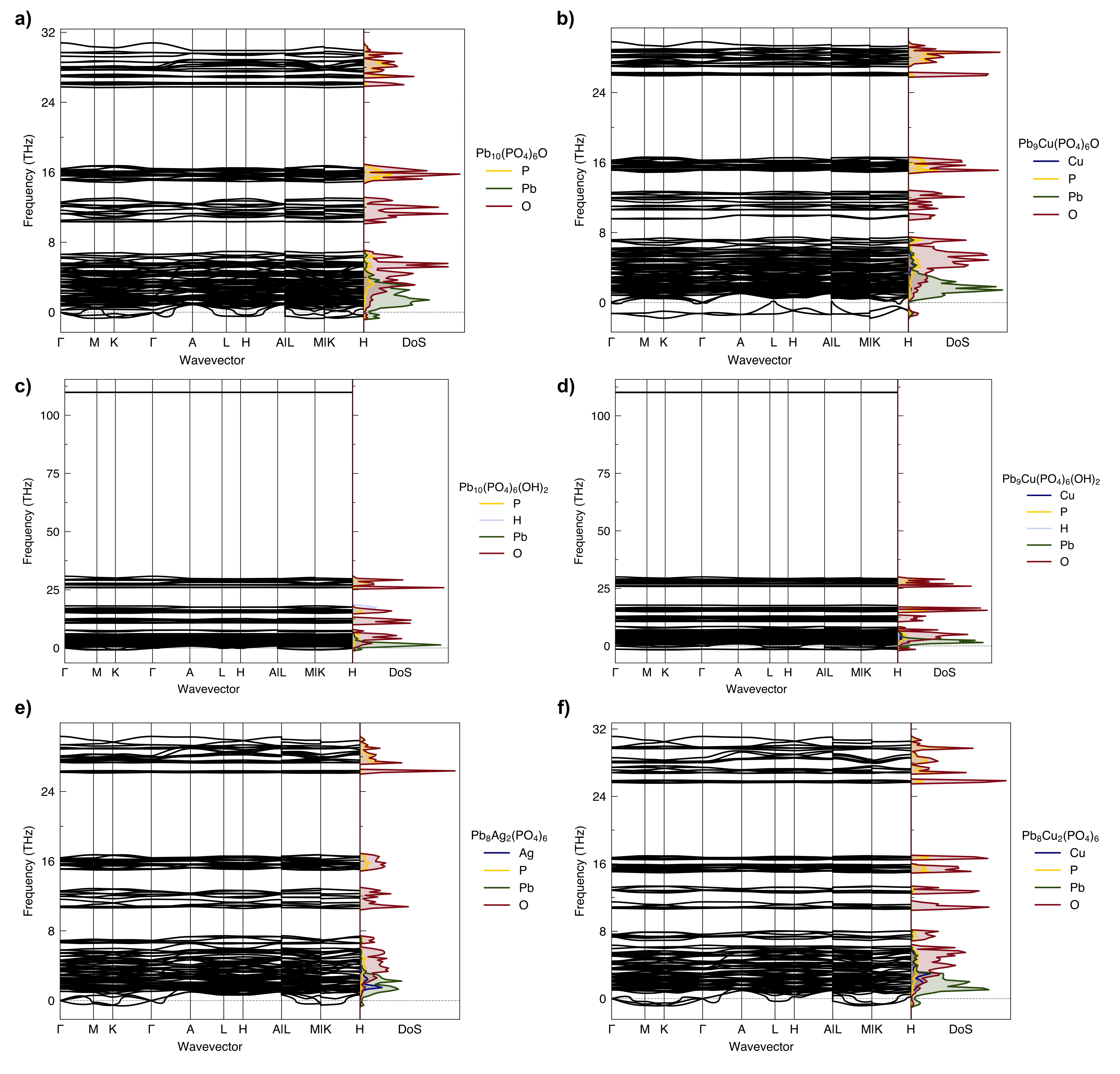}
\caption{In the main text, phonon band structures were truncated at 15 THz to allow for easier visualization of the imaginary phonon modes. Here, we present harmonic phonon band structures including all vibrational modes.}
\label{figs8}
\end{figure*}


\end{document}